\documentclass[12pt,a4paper]{article}
\usepackage[]{amsmath,amssymb}
\usepackage{graphicx}
\usepackage{geometry}
\usepackage{amssymb,epsfig,subfigure}
\usepackage{graphics,epsfig}
\usepackage{graphicx, color}
\usepackage{geometry}
\usepackage{diagbox}
\usepackage{float}
\usepackage{xcolor}
\usepackage[latin1]{inputenc}
\usepackage[T1]{fontenc}
\usepackage{amsmath}
\usepackage{amsfonts}
\usepackage{amssymb}
\usepackage{latexsym}
\usepackage[english]{babel}
\newcommand{\be}{\begin{equation}}
\newcommand{\ee}{\end{equation}}
\newcommand{\beq}{\begin{equation}}
\newcommand{\eeq}{\end{equation}}
\newcommand{\bea}{\begin{eqnarray}}
\newcommand{\eea}{\end{eqnarray}}

\def\be{\begin{equation}}
\def\ee{\end{equation}}
\def\ba{\begin{eqnarray}}
\def\ea{\end{eqnarray}}

\definecolor{princetonorange}{rgb}{1.0, 0.56, 0.0}
\definecolor{WildStrawberry}{rgb}{1.0, 0.26, 0.64}
\definecolor{rossocorsa}{rgb}{0.83, 0.0, 0.0}
\definecolor{navyblue}{rgb}{0.0, 0.0, 0.5}
\usepackage{hyperref}
\hypersetup{
    colorlinks,
    citecolor=rossocorsa,
    filecolor=navyblue,
    linkcolor=navyblue,
    urlcolor=navyblue
}

\begin{document}
\title{Instability of Universal Terms in the Entanglement Entropy}
\author{Marina Huerta\footnote{e-mail: marina.huerta@cab.cnea.gov.ar} , Guido van der Velde\footnote{e-mail: guido.vandervelde@ib.edu.ar }\\
{\sl Centro At\'omico Bariloche,
8400-S.C. de Bariloche, R\'{\i}o Negro, Argentina}}

\date{}

\maketitle
\begin{abstract}
 The role of symmetries in what concerns entanglement entropy has been extensively explored in the last years and revealed a profound connection with the quantum field theory's algebraic structure. Recently, it was found that some universal contributions to the entanglement entropy and mutual information may be non uniquely defined in theories with generalized symmetries. Here, we study this issue in detail in the particular case of the entanglement entropy of the Maxwell theory in $(2+1)$ dimensions for rotationally symmetric regions. In this setup, the problem can be dimensionally reduced to a half-line. We find that the only difference between the reduced problem for the Maxwell field and the reduced scalar free field stems from the Fourier angular $n=0$ mode. This simplification allows us to check explicitly the many issues that characterize models with broken global symmetries. Namely, we manifestly show that the additive algebras break Haag duality, and single out the non-local operators which are responsible for the failure of this property. More interestingly, we present concrete lattice realizations that confirm that the logarithmic "universal" term of the Maxwell entanglement entropy for disks depends on the details of the algebra assignation. This ambiguity hinders the identification of possible topological contributions characteristic of models with generalized symmetries and  tarnishes its universal character. We further calculate the Maxwell mutual information for two nearly complementary concentric disks. We obtain the expected universal contribution with a log-log dependence  and check that, unlike entropy, this is stable.  Accordingly, this supports mutual information as the appropriate probe to sense  additivity-duality breaking and the consequent universal topological contributions.

\end{abstract}

\section{Introduction}
\label{intro}
After a fruitful period of research, entanglement entropy (EE) has been shown to play a relevant role in the characterization of Quantum Field Theories (QFT). At present, it is well stated that the EE divergences structure has a strong geometric character and reveals universal features of the theory when choosing appropriate states and regions. The landscape in this regard is more subtle and vast than we could naively expect: depending on our theory and specific interest we have at disposal not only different choices of states and regions but also different nets of algebras, that is, different assignations of the operators available in the model to regions. In the last years, there has been a lot of progress in the understanding of this last point. 

As we could expect, recent investigations \cite{{Casini:2019kex},{Casini:2020rgj},{Casini:2021zgr}} reveal that the non-uniqueness of the assignation of a local algebra to a region can be, in general, understood as a source of ambiguities for the entanglement entropy. In this regard,  symmetries play a fundamental role in distinguishing apparent (ambiguities) from fundamental differences in the universal terms of the EE.
    The discussion about the EE ambiguities precedes this last issue. Historically,  the first source of ambiguities discussed in the literature is due to the regularization scheme dependency of the EE that reveals only some pieces of the EE are universal \cite{Calabrese:2004eu,Casini:2009sr,Klebanov:2011gs,Liu:2012eea, Grover:2012sp}. Later, it was noticed that also the particular choice of the operator content on the boundary, such as the addition of a center, a set of operators commuting with all the operators within the region, could affect the EE and modify the  universal terms \cite{Casini:2013rba,Casini:2014aia}. Within the framework of this discussion, the topological theories \cite{Kitaev:2005dm,Levin:2006zz} and the Maxwell field in $(3+1)$ dimensions are emblematic examples \cite{Buividovich:2008gq, Donnelly:2012st, Donnelly:2015hxa,Ghosh:2015iwa,Huang:2014pfa, Soni:2016ogt}. Regarding these two sources of ambiguities, one way to get rid of them but keeping track of the EE universal pieces of interest is to consider instead other information quantities, such as mutual information (MI) or more generally relative entropies, well defined in the QFT's context. More precisely, these quantities are finite and independent of the regularization scheme and chosen conveniently also contain the EE universal pieces. Moreover, relative entropies are also independent of how we choose the operator content on the boundary: the classical contribution to the MI due to the addition of a center located on the boundary studied in \cite{Casini:2014aia} does not survive the continuum limit, leaving the mutual information invariant. From this perspective and returning to the algebra-region assignation problem, it was shown in \cite{Casini:2019kex,Casini:2020rgj} that only in models where the duality or additivity property of the algebras is broken, there is a "genuine" physical, non-unique algebra choice, affecting universal pieces. In other words, the uniqueness of universal terms against a particular algebra choice depends on the failure of these properties which, in turn, depends on the completeness of the theory.

We will call a model incomplete when the algebra generated by the local degrees of freedom does not coincide with the maximal one compatible with causality. For example,  this is typically the case of a model associated with the neutral subalgebra that does not contain any charged operator. More generally, orbifolds belong to this family: they are obtained from a theory (complete) associated with an algebra $\cal{F}$ by retaining only the set of operators invariant under the action of a symmetry group $G$.  For these models, there is no way to satisfy duality and additivity at the same time. In restoring duality, for example, it is necessary the addition of nonlocal operators that break additivity and vice versa, if instead, we choose to restore additivity. Remarkably, additive algebras lead to different universal terms than those associated with dual algebras, even when these are read from well-defined relative entropies. The emergence of different universal terms notably distinguishes this scenario from the ones described previously. Unlike the ambiguities coming from regularization scheme prescriptions or boundary centers, the nonuniqueness of algebras assignations in incomplete models with global (or local) symmetries not only has a physical basement but an explicit manifestation revealed through topological contributions that alter the EE and MI universal terms.

In this article, we concretely illustrate the previous ideas for the particular case of the Maxwell field in $(2+1)$ dimensions.  This is a very appropriate stage for this purpose, because it not only admits a detailed and very accurate numerical treatment but also provides with a clear-cut manifestation of the algebra-region problem.

We follow the algebraic perspective of \cite{{Casini:2019kex},{Casini:2020rgj},{Casini:2021zgr}}, where these issues are addressed in great detail for models with global or local symmetries. 
The Maxwell field in $(2+1)$ dimensions is dual to the scalar theory, more precisely, to the subalgebra generated by the derivatives of the field. The subalgebra can be understood as the orbifold that results by quotienting the full algebra by the symmetry $\phi\rightarrow \phi+\text{const.}$. This model also exhibits spontaneous symmetry breaking (SSB) which, as we will see later, makes the definition of a regularized entropy in terms of MI more subtle or impossible. As extensively discussed in \cite{Casini:2019kex}, incomplete models with SSB present fundamental algebra region problems even for regions with  trivial topology. The case we consider here belongs to this class and according to the results in \cite{Casini:2019kex} the particular choice of the algebra affects non trivially the EE, naturally, but also the universal terms in the mutual information.

Here, we are particularly interested in the two following  results predicted in \cite{Casini:2019kex} concerning the Maxwell field and derived for models with SSB:
\be\label{log}
S_{scalar}-S_{Maxwell}=-\frac{1}{2}\log(R/\delta)
\ee
for the entropy of a disk of radius $R$ and $\delta$ the UV cutoff and 
\be\label{loglog}
I_{scalar}-I_{Maxwell}\sim\frac{1}{2}\log(\log(R/\epsilon))\,,
\ee
for the mutual information between  a disk of radius $R^-=R-\frac{\epsilon}{2}$ and the exterior of a disk $R^+=R+\frac{\epsilon}{2}$  separated by a distance $\epsilon$  in the almost complementary regions limit $\epsilon\rightarrow 0$.

The result (\ref{log}) was previously derived by means other than the algebraic approach, for instance, employing a direct wave function calculation \cite{Metlitski:2011pr} or the replica method \cite{Agon:2013iva}. Later, it was numerically confirmed in a square lattice for different choices of the boundary center \cite{Casini:2014aia}.

 It is important to note that, within the algebraic perspective, the differences in (\ref{log}) and (\ref{loglog}) have a concrete interpretation since they measure the change in universal terms for two  different algebra assignations. The scalar algebra contains the maximal set of operators (consistent with causality), while the Maxwell one contains only the symmetry invariant subset. It is interesting to note the unexpected difference in the dependence on $R$ in (\ref{log}) and (\ref{loglog}) which, as we will discuss later in detail, but anticipate here, is related to the additivity/duality tension for models with broken symmetry. Moreover and consistently with this last remark, we will show equation (\ref{log}), contrary to what happens for the MI difference in equation (\ref{loglog}), is not satisfied for all lattice regularizations. This tells us there is an instability affecting the entanglement entropy universal terms.  In other words, if equation (\ref{log}) depends on the regularization, then, it is simply non universal. This is one of the most important results in this article, which is organized as follows.

We start reviewing very briefly the algebraic perspective of the algebra-region problem for models with global symmetries. We introduce here definitions, properties, and results that will be useful for the rest of the discussion.
Then,  we focus on the Maxwell model. We start describing the model in the continuum and then its realization in a one-dimensional radial lattice. The circular symmetry allows us to dimensionally reduce the 
problem by integrating out the angular dependence. This reduction results in an infinite set of fields, one for each angular mode. As we are looking for differences between Maxwell theory and the full scalar, we center our attention in the zero mode, which is the only one that distinguishes them. We explore different subalgebra regularizations and show that Haag duality is manifestly broken. We calculate numerically $\Delta S=S_S-S_M$  in disks and remarkably find that only for some lattice realizations, all of them with the same continuum limit,  this difference is logarithmically divergent in the disk radius, as shown in (\ref{log}). The instability of $\Delta S$ reinforces the idea suggested in \cite{Casini:2019kex} that entropy differences might not be good enough to detect the presence of multiple (fundamental) algebraic assignments. This is in the reasoning line according to which entropies suffer ambiguities that could not be cured in general, even considering differences. The only reliable cases are the ones that stem from relative entropies. 
In this line, we know there is a fundamental multiplicity that affects the model that can be traced by calculating $\Delta I=I_S-I_M$. In fact, we numerically check this quantity for different subalgebra choices and find $\Delta I=1/2\log\log (R/\epsilon)$ in all cases, in perfect agreement with (\ref{loglog}). In all the calculations, it was helpful the consideration of the commutant algebras for the unbounded regions, not only for numerical calculation convenience but also to provide the numerics of physical interpretation. 

For pure states, when Haag duality holds,  we expect $S(V)=S(V')$, where $V'$ is the complementary region of $V$. In the present case, the commutant algebra is not simply the algebra sitting on the complementary region ($\mathcal{A}_{V'}\neq \mathcal{A'}_V$), but rather an algebra with extra non-local operators that implement the symmetry in $V$. In this regard, we find some novel results concerning unbounded twist operators, which enter the game when the entropy of unbounded regions, such as the exterior of a disk, is taken into account. In the present analysis, it is important to note that to obtain relevant information about the model in the continuum, our numerical results coming from calculations on a finite radial lattice must pass first through the IR limit of infinite lattice size and then the UV limit of null lattice spacing. The subtleties that arise in this respect are discussed in section \ref{results}.
Finally, we end the discussion with some concluding remarks.

\section{Algebras and Regions}\label{algebrasRegions}
For pure quantum states, the EE associated to a spatial region $V$ is a measure of the entanglement between the degrees of freedom (DOF) located inside and outside $V$ respectively. This definition obviously relies on the identification of the local operators belonging or not to $V$, which induces a partition of the complete Hilbert space as a tensor product ${\cal{H}}={{\cal{H}}}_V\otimes{{\cal{H}}}_{V^{\prime}}$, $V^{\prime}$ being the complementary region whose space-time points are spatially separated from those of $V$. The EE $S_V$ defined as the von Neumann entropy $S_V=-\text{tr}\rho_V\log\rho_V$ depends on the reduced density matrix $\rho_V$ which is in turn obtained from $\rho$ by  tracing over the DOF in $V^{\prime}$: $\rho_V=\text{tr}_{V^{\prime}}\rho$. When the state $\rho$ is pure, we have in general $S_V=S_{V^{\prime}}$.
In the language of algebras, this is related to the duality property of local algebras
\be 
{\cal{A}}_{V^{\prime}}={\cal{A}}^{\prime}_V
\ee
called Hagg duality \cite{Haag,Horuzhy,Bisognano:1975ih}. $\cal{A^{\prime}}$ is the commutant algebra, containing the operators that commute with operators in $\cal{A}$. On the other hand, for two causally complete regions $V_1$ and $V_2$ we expect local algebras to satisfy the additivity property
\be 
{\cal{A}}_{V_1}\vee {\cal{A}}_{V_2}={\cal{A}}_{V1\vee V2}
\ee
where $V1\vee V2=(V_1\cup V_2)^{\prime\prime}$ the smallest causally complete region containing the two.
It was found in \cite{Casini:2019kex,Casini:2020rgj} that these properties are not granted to be satisfied in models with an "incomplete" operator content. This is the case of models associated to a subalgebra $\cal{O}$ containing only the operators of the full algebra $\cal{F}$ that are invariant under the action of certain global symmetry group $G$. These are the orbifolds $\cal{F}/\cal{G}$.

Let us be more precise. Given a net, the additive algebra for a region $V$ can be constructed from those of $B\subset V$ as
\be
{\cal A}_{\textrm{add}}(V)= \bigvee_{B \,\textrm{is a ball}\,, \,B\subseteq V} {\cal A}(B)\,. 
\ee
This is the minimal algebra that contains all operators locally formed in $V$. Suppose that ${\cal A}_{\textrm{add}}(V)\subsetneq {\cal A}(V)$, then it is clear that we can have different nets with the same operator content of the full theory.

Among these possible algebra choices, we can identify the greatest one that can be assigned to $V$ and still satisfy causality. In turn, this must correspond to a minimal one assigned to $V'$,
\be
{\cal A}_{\textrm{max}}(V)= ({\cal A}_{\textrm{add}}(V'))'\,.
\ee
Evidently if ${\cal A}_{\textrm{add}}(V)\subsetneq {\cal A}_{\textrm{max}}(V)$ it follows that the additive net does not satisfy duality. In order to restore duality,  one can enlarge the additive net by adding non locally generated operators. In general this may be done in multiple ways. We will call such nets Haag-Dirac (HD) nets. Haag-Dirac nets satisfy duality
\be
{\cal A}_{\textrm{HD}}(V)=({\cal A}_{\textrm{HD}}(V'))'\,,
\ee
but by construction  will not satisfy in general additivity. This is the tension we refer to in the introduction that clearly cannot be avoided in incomplete theories. It is important to notice that for global pure states the entropy of an algebra ${\cal A}$ is equal to the one of its algebraic commutant ${\cal A}'$. The present discussion shows this does not imply an equality of entropies for complementary regions, except for a $HD$ net.    

Following the notation of \cite{Casini:2020rgj}, let us call $a \in {\cal A}_{\textrm{max}}(V)$ a collection of non locally generated operators in $V$ such that 
\be
{\cal A}_{\textrm{max}}(V)=({\cal A}_{\textrm{add}}(V'))' = {\cal A}_{\textrm{add}}(V)\vee \{a\}\,.
\label{a}
\ee
and $b \in {\cal A}_{\textrm{max}}(V')$ a set of operators non locally generated in $V'$ such that
\be
{\cal A}_{\textrm{max}}(V')=({\cal A}_{\textrm{add}}(V))' = {\cal A}_{\textrm{add}}(V')\vee \{b\}\,.
\label{b}
\ee
From their own definition, it is clear that the {\sl dual} sets of operators $\{a\}$ and $\{b\}$ cannot commute with each other. Moreover, their algebra can be interpreted as a generalized symmetry in the sense of \cite{Aharony:2013hda}, directly linked to the symmetries involved in the model. 

In summary, in constructing  Haag-Dirac nets ${\cal A}_{\textrm{HD}}(V)$ satisfying duality we have to resign some operators of ${\cal A}_{\textrm{max}}(V)$ or ${\cal A}_{\textrm{max}}(V')$: causality dictates we cannot take all possible operators  both for $V$ and $V'$. In fact, the assignation ${\cal{A}}_{\text{max}}(V)$ for all $V$ does not form a net.

In the present context, the relevance of the study of the properties of different possible nets relies in the fundamental consequences these have on the universal terms in the EE. This problem will be analyzed in depth for the Maxwell field in the following sections. 

\section{The model: Entropy and Local Algebras}
\label{themodel}

Maxwell's theory in  $ (2 + 1) $ dimensions is an interesting example of an incomplete theory  leading to the failure of additivity and/or duality. In three dimensions these problems manifest very neatly, since Maxwell is dual to the free scalar, through the identity
\begin{equation}\label{Maxwell_scalar}
\partial_{\mu}\phi=\frac{1}{2}\epsilon_{\mu\nu\rho}F^{\nu\rho}.
\end{equation}
More concretely, the model is that of the derivatives of the scalar field and is naturally invariant under the transformation $\phi\sim \phi +c$, for any constant shift $c$.  Thus, it  can be treated with the tools developed for theories with global symmetries. 
 In fact, the general case of gauge theories in three dimensions is particular in this sense.
We know that pure gauge theories in $d$ dimensions present algebra-region problems for regions with non trivial homotopy groups $\pi_1$ and $\pi_{d-3}$ \cite{Casini:2020rgj}. On the other hand, for global symmetries, this happens for regions with non trivial $\pi_0$ and $\pi_{d-2}$. These two notions coincide in $d=3$ and the Maxwell field is, in fact, an example of this duality between the generalized gauge symmetry and the  global one.

According to this, as a theory with global symmetries, we expect that there might be many algebra choices for a region $V$ with non trivial homotopy groups $\pi_0(V)$ or $\pi_1(V)$. For example, we could associate to a couple of disjoint disks the algebra generated by local operators at each disk, that is $\mathcal{A}_{\text{add}}=\mathcal{O}_1\vee\mathcal{O}_2$, or we could rather consider as well other neutral operators belonging to the model, that create a charge in a disk and the opposite one in the other, thus being non-local in the region of interest. In the language of section \ref{algebrasRegions}, this operator, called intertwiner, belongs to the class $a$ defined in (\ref{a}) and have the form $e^{i\lambda(\phi(1)-\phi(2))}$. The maximal algebra $\mathcal{A}_{\text{max}}=\mathcal{O}_1\vee\mathcal{O}_2\vee \{\mathcal{I}_{12}\}$ is dual to the additive algebra of the complement $\mathcal{A}_{\text{max}}(V)=(\mathcal{A}_{\text{add}}(V'))'$. Alternatively, the dual picture is based on the complementary region $V'$, and involves operators in the class $b$ of (\ref{b}). In this case we have $({\cal A}_{\textrm{add}}(V))' ={\cal A}_{\textrm{max}}(V')= {\cal A}_{\textrm{add}}(V')\vee \tau$ where $\tau$ is the twist operator defined as $\tau=\exp Q_V$, with $Q_V=\int_V j_0=\int_V \partial_0 \phi$ the symmetry charge in $V$. In the following sections, we will repeatedly use this last approach, choosing additive algebras for the regions of interest and in considering its commutant, dealing with enlarged algebras containing twist operators. We will explicitly show that twist operators (or in their dual version intertwiners) are responsible for the appearance of new topological contributions. Being unavoidable contributions with topological character, they lead to universal terms.
We repeat here for convenience what it is expected in $d=3$:
\be
\Delta S=S_{scalar}-S_{Maxwell}=-\frac{1}{2}\log(R/\delta) \nonumber
\ee
for the entropy of a disk of radius $R$, with $\delta$ the UV cutoff, and
\be
\Delta I=I_{scalar}-I_{Maxwell}\sim\frac{1}{2}\log(\log(R/\epsilon))\nonumber
\ee
for the mutual information between  a disk of radius $R^-=R-\frac{\epsilon}{2}$ and the exterior of a disk $R^+=R+\frac{\epsilon}{2}$  separated by a distance $\epsilon \rightarrow 0$. 

Although we will not discuss here the details of the derivation of (\ref{log}) and (\ref{loglog}) in \cite{Casini:2019kex,Casini:2020rgj}, in order to contextualize the results, we emphasize that these are based on the identifications $\Delta S(R)=S_{\mathcal{ F}}(\omega\vert\omega\circ E)$ and $\Delta I(1,2)=S_{\mathcal{ F}}(\omega_{1}\vert\omega_{1}\circ E_{1})+S_{\mathcal{ F}}(\omega_{2}\vert\omega_{2}\circ E_{2})-S_{\mathcal{ F}}(\omega_{12}\vert\omega_{12}\circ E_{12})$. The relative entropy $S_{\mathcal{ F}}(\omega\vert\omega\circ E)$ defined on the full algebra $\mathcal{F}$ of the scalar, measures the "distance" between the state $\omega$ in the full model (here the vacuum for the free scalar) and the state $\omega\circ E$ which is the vacuum where the non local operators have been projected  to zero expectation values within the region (here a disk). The conditional expectation $E$ is the one in charge of this projection. Basically, $E$ acts on the elements of the full algebra selecting the part invariant under the symmetry group. The same generalizes to $S_{\mathcal{ F}}(\omega_{12}\vert\omega_{12}\circ E_{12})$, defined on the full algebra for two disconnected regions $1,2$ (here the disk of radius $R^-$ and the complement of a disk of radius $R^+$). The fact of having been able to express the differences $\Delta S$ and $\Delta I$ in terms of relative entropies simplifies enormously the discussion due to the properties of relative entropies: well defined in the continuum, ordered by inclusion, entropic certainty and uncertainty relations. In fact, we remark that  equation (\ref{log}) becomes meaningful only through the identification with a relative entropy, that in the present model is divergent. As we will discuss later and anticipated in the Introduction, this is not always the case: using the radial lattice regularization scheme, there are some choices of local algebras for the Maxwell field whose entropies do not satisfy $S_{\mathcal{O}}(\omega) \longleftrightarrow S_{\mathcal{F}}(\omega \circ E)$, leading to  $\Delta S \neq S_{\mathcal{ F}}(\omega\vert\omega\circ E)=S_{\mathcal{F}}(\omega)-S_{\mathcal{ F}}(\omega\circ E)$ and a consequent violation of equation (\ref{log}). For equation (\ref{loglog}) instead, involving MI, already well defined in the continuum no matter the regularization scheme we use, these concerns disappear, promoting the latter as a more reliable test for duality/additivity breaking.

We stress the fact that our model has SSB with an infinite symmetry breaking parameter. This is what makes divergent the relative entropy $S_{\mathcal{ F}}(\omega\vert\omega\circ E)$ in (\ref{log}) for regions with trivial topology.
\subsection{Conformal invariance}
We end by noting that Maxwell's theory in three dimensions is incomplete in a broader sense than the one discussed above. As explained in \cite{El-Showk:2011xbs}, Maxwell theory is scale invariant but not conformal invariant. This is because the gauge invariant operator with lowest scaling dimension, namely the field strength tensor $F_{\mu\nu}$, is not a primary. By the duality relation (\ref{Maxwell_scalar}) it becomes evident that it is a descendant of the scalar field, which does not belong to the Maxwell theory. In other words, we must complete the theory to the free scalar model if we want to save conformal invariance.

\begin{figure}[t]
\begin{center}  
\includegraphics[width=0.55\textwidth]{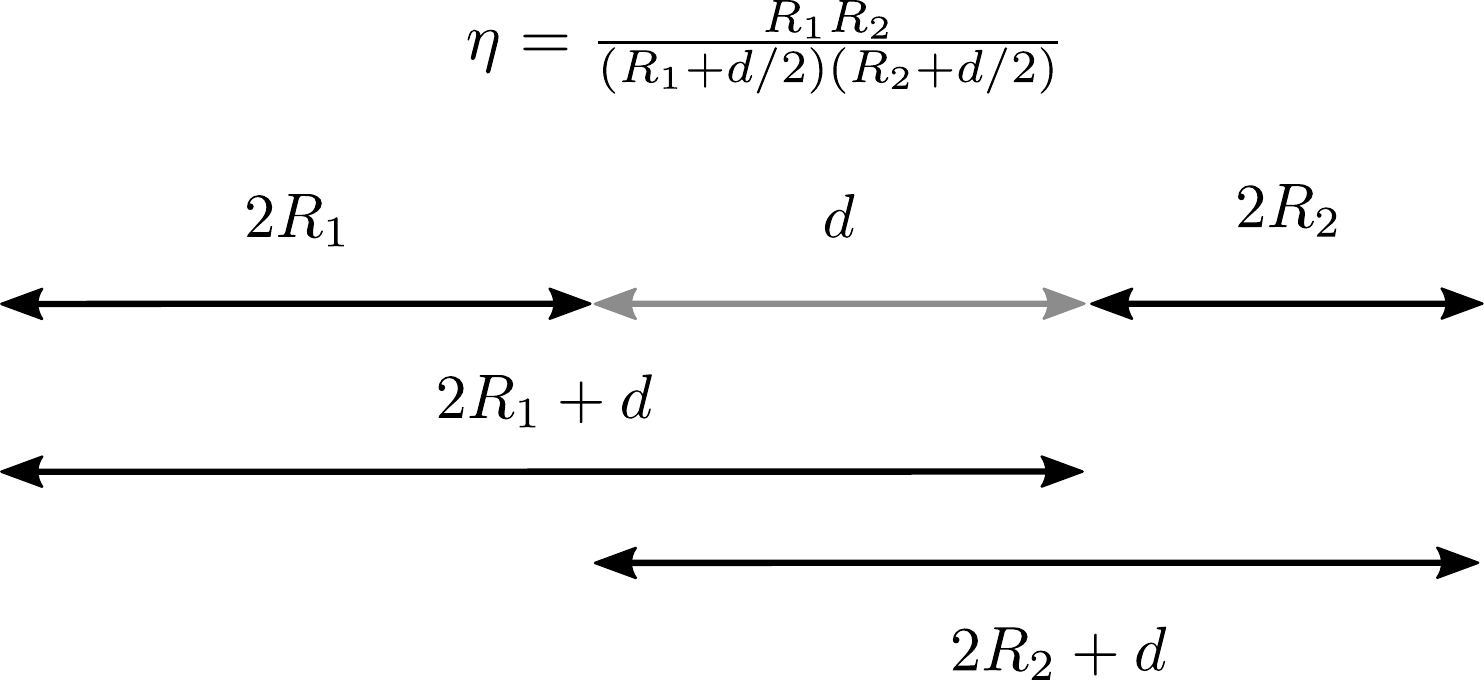}
\caption{Cross ratio for four points. $R_1$ and $R_2$ are the disk radii, whereas $d$ represents the separation between the disks.}
\label{figeta}
\end{center}  
\end{figure}

\noindent
A quantum information probe for the lack of conformal invariance of our model is mutual information.
For two non intersecting regions (here disks) $R_1$ and $R_2$, separated by a distance $d$, mutual information is
\begin{equation}
I(R_1,R_2,d)=S(R_1)+S(R_2)-S(R_1\cup R_2)
\end{equation}
\noindent
For conformal theories and non intersecting regions completely  characterized by four points, such as two separated disks, mutual information must be a function of the cross ratio $\eta=\frac{(x_2-x_1)(x_4-x_3)}{(x_3-x_1)(x_4-x_2)}$, or in terms of the disks' radii (figure (\ref{figeta}))
\begin{equation}
\eta=\frac{R_1R_2}{(R_1+d/2)(R_2+d/2)}
\end{equation}

\begin{figure}[t]
\centering
\subfigure{\includegraphics[width=0.48\linewidth]{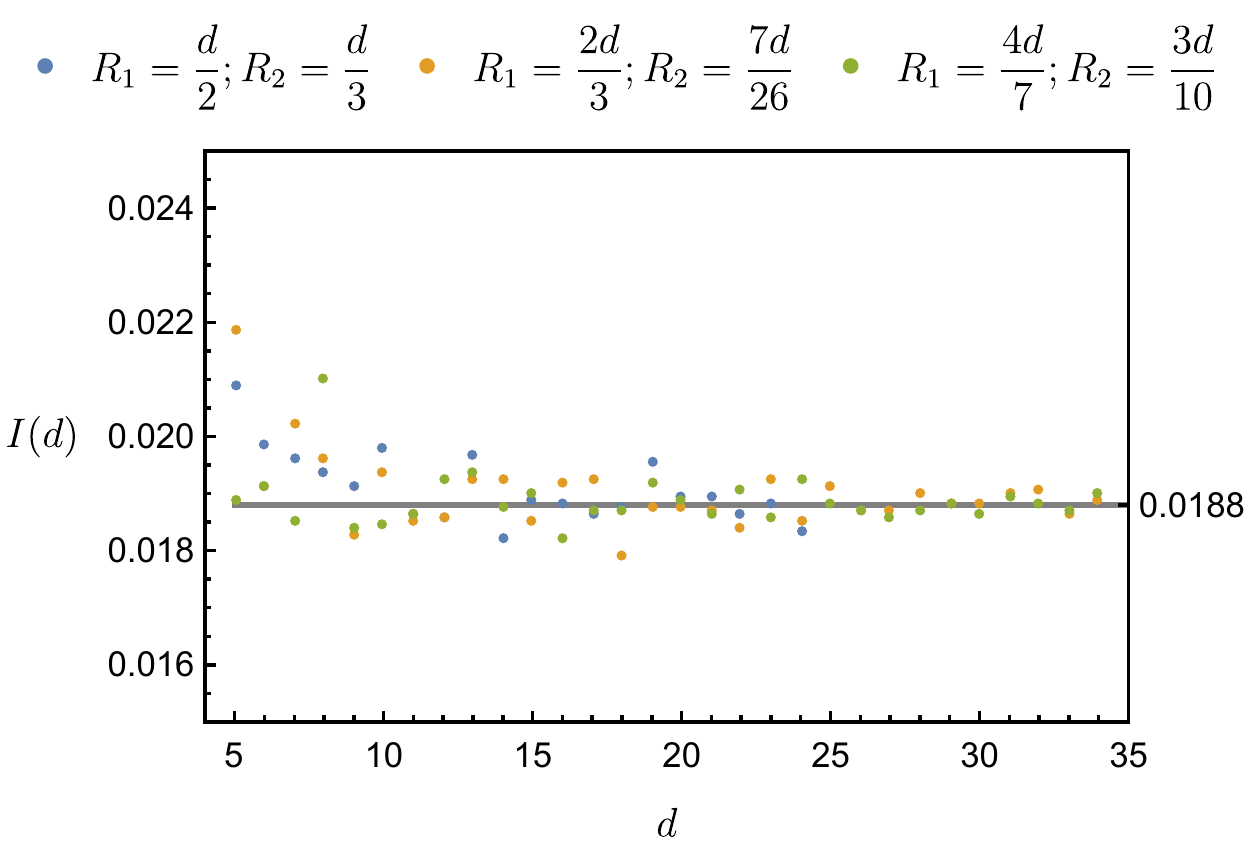}}
\hfill
\subfigure{\includegraphics[width=0.47\linewidth]{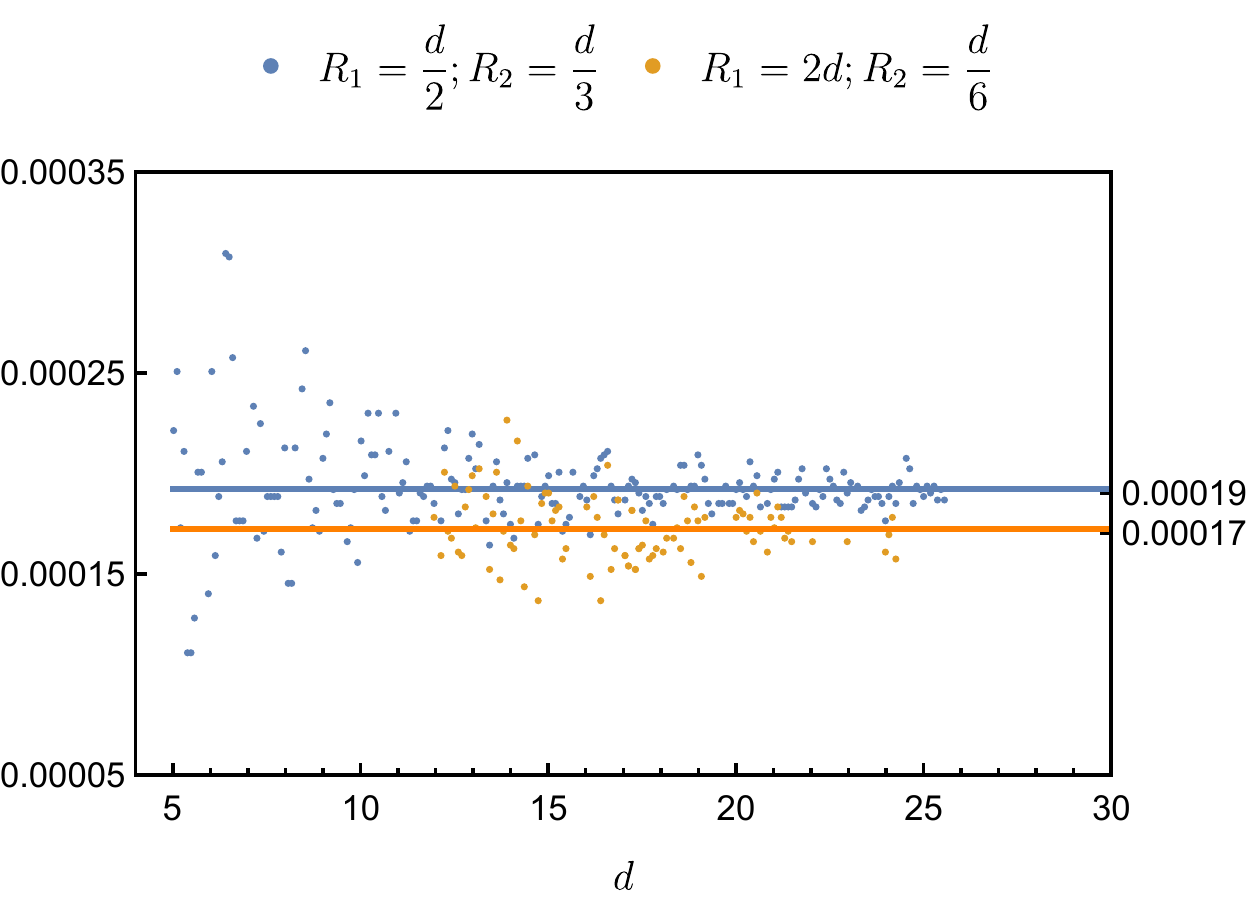}}
\caption{Mutual information vs. separation $d$ between disks. We define $\eta=\frac{R_1 R_2}{(R_1+d/2)(R_2+d/2)}$. At small $d$, where the size of the disks is comparable to the lattice site (a unit), mutual information oscillates. At large $d$, it converges to an asymptotic value, which represents the continuum limit. Left panel: scalar theory. Right panel: Maxwell theory.}
\label{mutual_eta}
\end{figure}

\noindent
 Thus, we compute mutual information in the square lattice, for disks of radius $R_1$ and $R_2$, separated by a distance $d$, and compare the results for different configurations with fixed $\eta=1/5$. See figure (\ref{mutual_eta}). The details of the numerical calculation can be found in \cite{Casini:2014aia} where the authors studied the Maxwell's EE for different geometries, circles included, in a two dimensional square lattice.

\noindent

As expected, in the full scalar theory mutual information converges to the same value for all the configurations, as long as the cross ratio $\eta$ is preserved. On the contrary, when we consider just the Maxwell subalgebra mutual information is not constant even at fixed $\eta$, which reflects the breaking of conformal symmetry.

\section{Scalar and Maxwell fields in $(2+1)$ dimensions: The Disk Entanglement Entropy}
In this section we review the necessary ingredients to calculate the entanglement entropy of the scalar and Maxwell fields in rotationally invariant regions. Taking advantage of this symmetry, we dimensionally reduce each system to a half line in the radial direction. We then regularize the degrees of freedom in a one dimensional lattice, and explain how to numerically compute the entanglement entropy from the lattice correlators and commutation relations.
\subsection{Modes decomposition}

The massless scalar field Hamiltonian in $(2+1)$ dimensions is

\begin{equation}
H= \frac{1}{2}\int{d^{2}x\lbrace \pi ^{2} + \left( \partial_{i} \phi\right)^{2}\rbrace}\,\,,\,\,i=1,2\,.
\end{equation}

As we are interested in the disk EE,  we find convenient the use of polar coordinates $r,\theta$. In these coordinates, the fields admit the following Fourier expansion

\begin{equation} \label{fourier}
\phi\left(r,\theta\right)=  \frac{1}{\sqrt{2\pi}} \sum_{n} \phi_{n}\left(r\right)e^{in\theta}\,,
\end{equation}
\begin{equation} \label{ec3}
\phi_{n}\left(r\right)=  \frac{1}{\sqrt{2\pi}} \int d\theta\, \phi\left(r,\theta\right)e^{-in\theta}\,,
\end{equation}

with analogous expressions for $\pi$. Besides, rescaling fields as

\begin{equation} \label{ec4}
\widetilde{\phi}_n=  \sqrt{r}\, \phi_n \,,
\end{equation}
\begin{equation} \label{ec5}
\widetilde{\pi}_n=  \sqrt{r} \,\pi_n \,,
\end{equation}
we recover standard commutation relations for the radial fields

\begin{equation}
\left[ \widetilde\phi_{n}\left(r\right),{\widetilde\pi_{n'}}^{\dagger}\left(r'\right)\right]=i\delta_{n,n'}\delta\left(r-r'\right)\,.
\end{equation}

In these coordinates, the reduced Hamiltonian that results by integrating the angular variable is

\begin{equation} \label{hamiltonianoescalar}
H= \sum_{n} \frac{1}{2}\int_{0}^{\infty}{dr\left\{ \widetilde{\pi}_{n} ^{2} + r\left( \frac{\partial}{\partial r} \left(\frac{\widetilde{\phi}_{n}}{\sqrt{r}}\right)\right)^{2}+\left(\frac{n}{r}\right)^{2}\widetilde{\phi}_{n}^{2}\right\}}\,.
\end{equation}

We note $H_0$ (equation (\ref{hamiltonianoescalar}) with $n=0$), is not equivalent to the Hamiltonian of a $(1+1)$ dimensional massless scalar.  More precisely, analyzing  the second term of the above Hamiltonian we find

\begin{equation}
r\left[\frac{\partial}{\partial r}\left(\frac{\phi}{\sqrt{r}}\right)\right]^2=r\left[\frac{\phi'}{\sqrt{r}}-\frac{\phi}{2r^{3/2}}\right]^2=r\left[\frac{\left(\phi'\right)^2}{r}+\frac{\left(\phi\right)^2}{4r^{3}}-\frac{\phi'\phi}{r^2}\right]=\left(\phi'\right)^2+\frac{\left(\phi\right)^2}{4r^{2}}-\frac{\phi'\phi}{r}
\end{equation} 

where $(\phi')^2$ in the last identity coincides with the one dimensional scalar and 

 \begin{equation}
\frac{\left(\phi\right)^2}{4r^{2}}-\frac{\phi'\phi}{r}=-\left(\frac{\phi ^2}{2r}\right)'-\frac{\phi ^2}{4r^2}\,,
\end{equation}
gives an extra contribution $\frac{\phi ^2}{4r^2}$.
Moreover, in general  for $n\neq 0$, the Hamiltonian $H_n$ corresponds to a one dimensional scalar with a quadratic contribution $\frac{\phi^2}{r^2}\left(n^2-\frac{1}{4}\right)$.

On the other hand, the Hamiltonian of the Maxwell theory is usually written in terms of the electric and magnetic physical fields $E$ and $B$

\begin{equation}\label{hamiltoniano}
\begin{split}
H&= \frac{1}{2}\int{d^{2}x\left( B^2 +E^2\right)}\\
&= \sum_{n}\frac{1}{2}\int_{0}^{\infty} r dr \left\{ B_{n}^{2} + \left(E_{n}^{r}\right)^2 + \left(E_{n}^{\theta}\right)^{2}\right\}\,.
\end{split}
\end{equation}

Equation (\ref{Maxwell_scalar}) can be inverted to yield, in polar coordinates\footnote{Note that the $1/r$ factor that relates $F^{\theta t}$ with $E^{\theta}$ comes from the fact that the latter is the component of the electric field in the $\hat{\theta}$ direction, rather than an angular component in the tangent space spanned by $\left\lbrace\frac{\partial}{\partial r},\frac{\partial}{\partial \theta}\right\rbrace$  },
\begin{equation}
\begin{split}
F^{r\theta}&=-\frac{1}{r}\partial_0\phi=\frac{1}{r}B\\
F^{r t}&=\frac{1}{r}\partial_{\theta}\phi=E^r\\
F^{\theta t}&=-\frac{1}{r}\partial_r \phi=\frac{1}{r} E^{\theta}.
\end{split}
\end{equation}
This leads to the following identities between the Fourier modes of the electric/magnetic and the scalar fields
\begin{equation}\label{Maxwell_scalar_modes}
\begin{split}
B_n &=-\frac{1}{\sqrt{r}}\partial_0\tilde{\phi}_n\\
E_n^r&=\frac{i n}{r}\phi_n=\frac{i n}{r^{3/2}}\tilde{\phi}_n\\
E_n^{\theta}&=-\partial_r \phi_n=-\partial_r \left(\frac{\tilde{\phi}_n}{\sqrt{r}}\right).
\end{split}
\end{equation}
Substituting (\ref{Maxwell_scalar_modes}) into (\ref{hamiltoniano}), we recover the scalar reduced Hamiltonian, as expected. However, eq. (\ref{Maxwell_scalar_modes}) also shows that when $n\neq0$ there is a one to one correspondence between the scalar mode $\tilde{\phi}_n$ and $E_n^r$. In other words, not only the reduced radial theories coincide, but also their operator content. This is because the transformation $\phi\rightarrow \phi +c$ translates into $\phi_n\rightarrow\phi_n+\sqrt{2\pi}c\delta_{n,0}$ in Fourier space leaving the $n>0$ modes invariant. 

Instead, the zero mode satisfies $E_{n=0} ^r=0$, so the Maxwell field is equivalent to the derivative of the scalar field. In this case, if we define $\tilde{B_0}=\sqrt{r}B_0$, the Hamiltonian reads

\begin{equation}
H_{0}= \frac{1}{2}\int_{0}^{\infty}{dr\left\{ \widetilde{B}_{0} ^{2} +r \left(E_{0}^{\theta}\right)^{2}\right\}}
\end{equation} 

and the commutation relation between the gauge invariant degrees of freedom is
\begin{eqnarray}
\left[E_0^{\theta}(r),\tilde{B}_0(r')\right]&=&\left[\partial_r\left(\frac{\tilde{\phi}_0(r)}{\sqrt{r}}\right),{\tilde{\pi}}_0(r')\right]\\
&=& i\partial_r\left(\frac{\delta(r-r')}{\sqrt{r}}\right)\,.\label{corr_M}
\end{eqnarray}

\subsection{Numerics}
\label{numerics}
In order to make numerical computations we consider a collection $\{\phi_i,\pi_i\}$, each living in a lattice site $i$ with Hamiltonian
\begin{equation}\label{lattice_hamiltonian}
H=\sum_{ij}{\pi_i}^2+\phi_i K_{ij}\phi_j,
\end{equation}
satisfying
\begin{equation}
\left[\phi_i,\pi_j\right]=iC_{ij},
\end{equation}
which generalizes the case of canonically conjugate variables, where $C_{ij}=\delta_{ij}$.

For gaussian models the reduced density matrix $\rho_V$ of the vacuum in a given region $V$ can be expressed in terms of the correlators of the theory \cite{Casini:2009sr}. In turn, these can be read from the Hamiltonian (\ref{lattice_hamiltonian}) 
\begin{equation}
X_{ij}\equiv \langle\phi_i\phi_j\rangle=\frac{1}{2}K_{ij}^{-1/2}\, , \quad P_{ij}=\langle\pi_i\pi_j\rangle=\frac{1}{2}K_{ij}^{1/2}.
\end{equation}
When the algebra associated to $V$ has no center, or equivalently when ${C\rvert}_V$ is invertible, the entanglement entropy is \cite{Casini:2009sr}
\begin{eqnarray}
S(V)&=&\text{Tr}\left[\left(\Theta +\frac{1}{2}\right)\log{\left(\Theta+\frac{1}{2}\right)}-\left(\Theta -\frac{1}{2}\right)\log{\left(\Theta-\frac{1}{2}\right)}\right]\, \label{vonNeumann},\\
\Theta &=&\sqrt{C_V^{-1}X_V {(C_V^T)}^{-1} . P_V}
\end{eqnarray}
where the sub index $V$ means that we restrict the matrices to the degrees of freedom in $V$.

Conversely, if we choose an algebra generated by all the $\pi_j$ operators in $V$ (with $j\in \lbrace 1,...,n\rbrace$), but only a subset of $\phi_i$ (with $i\in B=\lbrace 1,...,k\rbrace$), such that the operators $\pi_{\ell}, \ell\in A=\lbrace k+1,...,n\rbrace$ span a center, the EE is the sum of two terms
\begin{equation}
S(V)=S_Q (V)+H(A).
\label{vonNeumann2}
\end{equation}
The first term is a quantum contribution, equal to (\ref{vonNeumann}), but defining instead \cite{Casini:2014aia}
\begin{eqnarray}
\Theta &=& \sqrt{\widetilde{X}\widetilde{P}}\, ,\\
\widetilde{X} &=& C_B^{-1}X_B {(C_B^T)}^{-1}\, ,\\
\widetilde{P}&=&\left[{P_V}^{-1}\rvert_B\right]^{-1}.
\end{eqnarray}
The second term is a classical contribution, that is the Shannon entropy due to the operators in $A$
\begin{equation}\label{Shannon}
H(A)=\frac{1}{2}\text{Tr}\left(1+\log{(2\pi P_A})\right).
\end{equation}
This is an ambiguous quantity that depends on the normalization of the operators in $A$, which is not fixed by the commutation relations. However, relative entropies and mutual information are independent of this choice, hence being well defined measures.

In summary, once the lattice Hamiltonian of the system is identified and an algebra of operators is assigned to a region, the EE can be calculated numerically, as explained above. In particular for the scalar and Maxwell cases, from the dimensionally reduced Hamiltonian (\ref{hamiltonianoescalar}), we get for each scalar mode $n$
\begin{eqnarray}
K^n_{1,1} &=& \frac{3}{2} + n^2;\\
 K^n_{i, i} &=& \frac{n^2}{i^2} + 2; \,\,\,i= 2, ... ,m\\
 K^n_{i, i + 1} &=& K_{i+1,i}=\frac{-(i + \frac{1}{2})}{\sqrt{i (i + 1)}};\,\,\,i= 1, ... m-1
\end{eqnarray}
where $m$ is the lattice total size, and
\begin{equation}
C_{ij}=\delta_{ij},
\end{equation}
since $\widetilde{\phi}^n_i$ and $\widetilde{\pi}^n_j$ are canonically conjugated variables. The total EE is a sum over the EE of each independent mode.

The $n>0$ part of the Maxwell EE is exactly the same as the scalar's, because both the Hamiltonians and the algebras coincide. All the difference is in the $n=0$ contribution. Nevertheless, we can profit from the identification of (\ref{Maxwell_scalar_modes}) to compute the correlators of the gauge invariant operators $\phi_{M i}\equiv E_{0i}^{\theta}=\frac{\widetilde{\phi}_{i+1}}{\sqrt{i+1}}-\frac{\widetilde{\phi}_{i}}{\sqrt{i}}$ as
\begin{equation}
X_{M i j}\equiv\langle \phi_{M i}\phi_{M j}\rangle=\frac{\widetilde{X}_{i+1 j+1}}{\sqrt{(i+1)(j+1)}}+\frac{\widetilde{X}_{ij}}{\sqrt{ij}}-\frac{\widetilde{X}_{ij+1}}{\sqrt{i(j+1)}}-\frac{\widetilde{X}_{i+1j}}{\sqrt{(i+1)j}}. \label{corM}
\end{equation}
Additionally, from (\ref{corr_M}) we can read
\begin{equation}
C_{ij}\equiv -i\left[\phi_{M i},\pi_{M j}\right]=\frac{\delta_{i+1,j}}{\sqrt{i+1}}-\frac{\delta_{i,j}}{\sqrt{i}}.
\end{equation}
   
\section{Universal terms: Analysis and Results}
\label{results}
\subsection{Maxwell and Scalar algebras in the radial lattice}
\label{Maxwell algebras in the radial lattice}
\begin{figure}[t]
\begin{center} 
\includegraphics[width=0.55\textwidth]{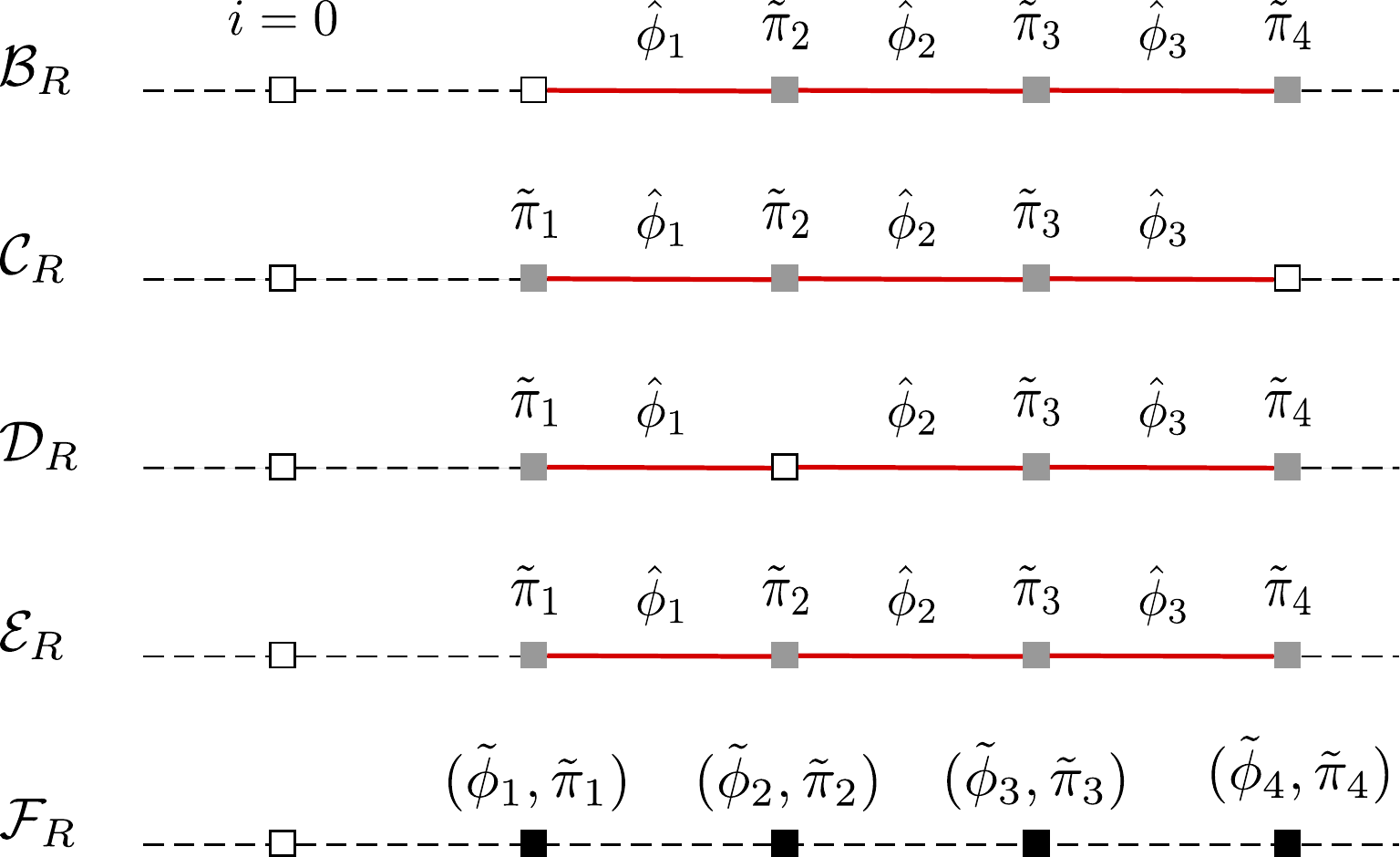}
\caption{Radial lattice: Link operators in red ${\phi_M}_i=\widehat{\phi}_i=\frac{\widetilde{\phi}_{i+1}}{\sqrt{i+1}}-\frac{\widetilde{\phi}_i}{\sqrt{i}}$, and momentum operators attached to the lattice vertices ${\pi_M}_i=\widehat\pi_{i}=\widetilde\pi_i$. The first four panels correspond  (from top to bottom)  to Maxwell subalgebras $\mathcal{B}_R$, $\mathcal{C}_R$, $\mathcal{D}_R$ and $\mathcal{E}_R$. Fifth panel corresponds to the full scalar algebra $\mathcal{F}_R$.}
\label{radiallattice}
\end{center}  
\end{figure}

Based on the observation  that the algebra-region assignation, in general non unique, affects dramatically the corresponding entanglement entropy, it is mandatory to explore this issue for the reduced Maxwell theory in the radial lattice. Here, intervals  represent spherically symmetric regions, such as a disk when the intervals are connected to the origin. In that case, the most natural choice for the Maxwell subalgebra would be to include all the momentum operators $\pi_M$ at the disk sites, as well as the links $\phi_M\equiv \hat{\phi}$ inside, depicted in red in figure (\ref{radiallattice}). This algebra has a center and its entropy is not purely quantum. Another possibility, that averts this issue, is to take one momentum out of the net, so that the number of conjugate operators gets balanced. One would be tempted to anticipate that in the continuum limit, even knowing the entropy is for sure sensitive to what the site with a missing momentum is, the universal part will not.  We will see later that this is not the case:  entropy differences as (\ref{log}) are sensitive to different choices and the identification with a relative entropy is not guaranteed for every lattice realization.

Let us introduce some definitions that will be used throughout the rest of the discussion.
In the radial lattice, consider a disk of radius $R$ corresponding to the interval of length $i_{max}$ attached to the origin, with $R= (i_{max}+\frac{1}{2}) \delta$ and $\delta$ the unit lattice spacing chosen here $\delta=1$. The full scalar algebra $\mathcal{F}_R$ associated to this region is chosen to be
\be
\mathcal{F}_R=\{\widetilde\phi_1,..,\widetilde\phi_{i_{max}} ,\widetilde\pi_1,..,\widetilde\pi_{i_{max}}\}
\ee
 as shown in the fifth panel of figure (\ref{radiallattice}). This is, each lattice point has a pair of canonically conjugated variables $(\widetilde\phi_i, \widetilde\pi_i)$.

On the other hand, the Maxwell model contains link operators ${\phi_M}_i\equiv\widehat{\phi}_i=\frac{\widetilde{\phi}_{i+1}}{\sqrt{i+1}}-\frac{\widetilde{\phi}_i}{\sqrt{i}}$, and momentum operators ${\pi_M}_i\equiv\widehat\pi_{i}=\widetilde\pi_i$. 

The first four panels in (\ref{radiallattice}), represent different subalgebra choices for a disk of radius $R=4+1/2$.
 We will call Maxwell subalgebras $\mathcal{B}_V$, $\mathcal{C}_V$ and $\mathcal{D}_V$ associated to a general one component region $V$, the nets with a momentum operator $\pi_k$ taken out from the first, the last, and a middle site respectively. On the other hand, we call subalgebra $\mathcal{E}_V$ that with all momentum operators in $V$. 
 \noindent
 
Note that the first three $\mathcal{B}$, $\mathcal{C}$ and $\mathcal{D}$, correspond to algebras without center. Concretely, the  operator content in these cases is
\begin{eqnarray}
{\cal{B}}_R&=&\{\widehat\phi_1,..,\widehat\phi_{(i_{max}-1)}, \widehat\pi_2,..,\widehat\pi_{i_{max}} \} \label{br}\\ 
{\cal{C}}_R&=&\{\widehat\phi_1,..,\widehat\phi_{(i_{max}-1)},\widehat\pi_1,..,\widehat\pi_{(i_{max}-1)}\}\label{cr}\\
{\cal{D}}_R&=&\{\widehat\phi_1,..,\widehat\phi_{(i_{max-1})},\widehat\pi_1,...,\widehat\pi_{k-1},\widehat\pi_{k+1}...,\widehat\pi_{i_{max}}\}
\end{eqnarray} 
 with balanced number of link and momentum operators in all three cases.
 Finally, the fourth panel represents the orbifold subalgebra $\cal{E}$ 
 
\begin{equation}
\mathcal{E}_R=\{\widehat\phi_1,..,\widehat\phi_{(i_{max}-1)}, \widehat\pi_1,..,\widehat\pi_{i_{max}} \}
\label{er}
\end{equation} 
\noindent
  It is easy to see that the subalgebra $\cal{E}$ has a center, this is, there is an operator $\sum_{i=1}^{i_{max}}\sqrt{i}\widehat\pi_{i}$ that commutes with all the operators in the region.

We will measure the relevance of the different choices in terms of how these affect the universal topological contributions in the entropy and mutual information differences (\ref{log}) and (\ref{loglog}). This will be the main subject of the following sections.
\subsection{Scalar in the disk}
Although the EE of the full scalar has no logarithmic term in odd dimensions, each mode in the Fourier decomposition contributes with a logarithmic piece, naturally expected in the dimensionally reduced model. The cancellation of the logarithmic pieces occurs only when summing over the total number of (infinite) modes. For instance, for the $n=0$ mode, which is the relevant one in the comparison with the Maxwell EE, we get
\begin{equation}\label{clog_scalar}
c_{\text{log}}^S=0.1666551\sim \frac{1}{6}.
\end{equation}
We find that this coefficient is completely stable and independent on the infrared cutoff of the lattice. We also note that this coincides with the logarithmic coefficient for the one dimensional scalar on the half line, as it should.

Nevertheless, the whole sum of modes reproduces the linear behavior
\begin{eqnarray}
S&=&c_1 R-c_0\, ,\label{F}\\
c_0&=&0.062811\, ,
\end{eqnarray}
where $c_0$ is the universal constant F-term \cite{Casini:2015woa,Bueno:2021fxb}. 
This serves us as second cross check for the $n=0$ result.

\begin{figure}[t]\label{F-term}
\begin{center}  
\includegraphics[width=0.55\textwidth]{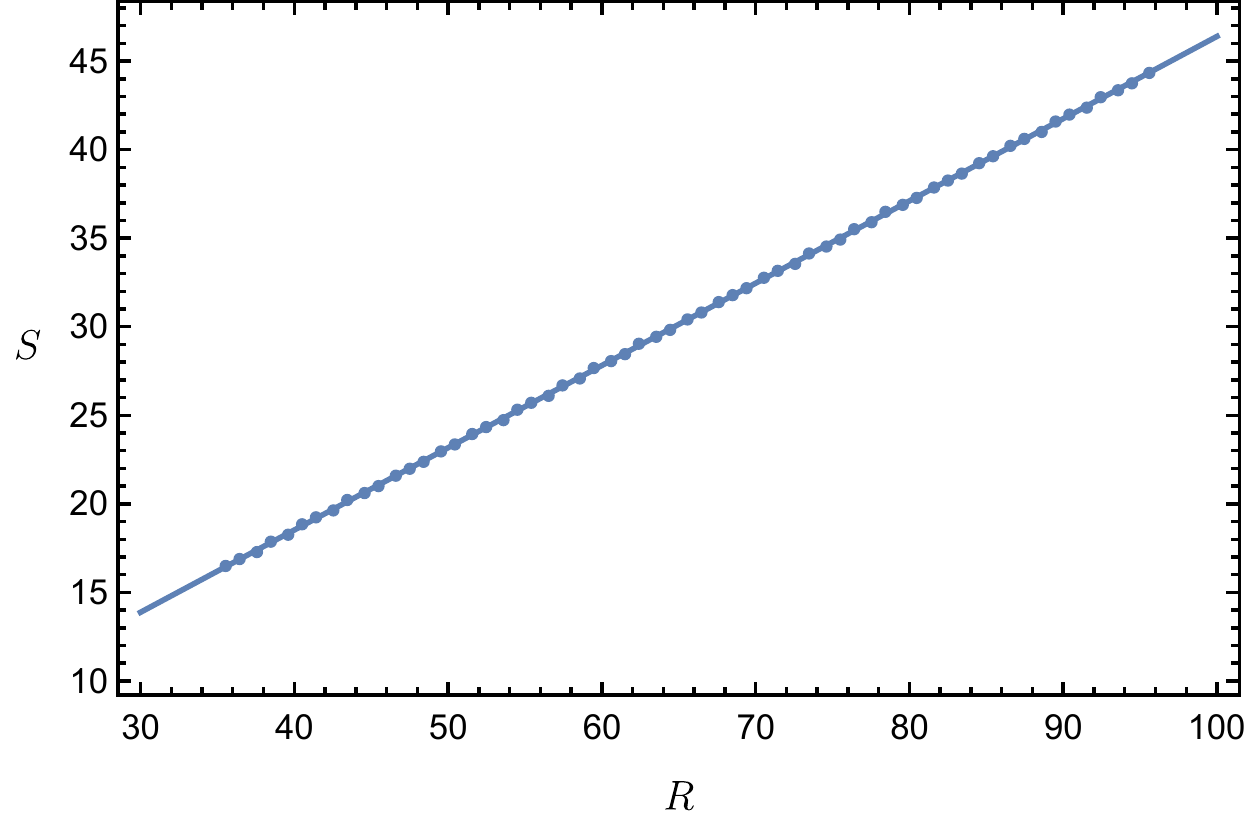}
\caption{Disk scalar entropy as a function of the radius $R$. The curve corresponds to the linear fit $S(R)\sim -0.062811+0.464578 R$.}
\label{figc0}
\end{center}  
\end{figure}

In (\ref{F}), the EE was computed exactly for every mode up to $n_{\text{max}}=2000$. Large $n$ corrections where taken into account by fitting the EE for the set $n=\lbrace 1500, 2000, 2500, 3000, 3500\rbrace$, with the function
\begin{equation}
S_n=\frac{c_a}{n^4}+\frac{ c_b}{n^4} \log[n]+ \frac{c_c}{n^6}+\frac{c_d}{n^6 }\log[n]+\frac{c_e}{n^8}+ \frac{c_f}{n^8} \log[n]\, ,
\end{equation}
and then summing over the best fit $\sum_{n=2001}^{\infty}S_n$. The procedure was repeated for lattice sizes $m=400,500,600,700,800,900$, and later the infrared limit was taken. The results are plotted in figure (\ref{figc0}).

\subsection{Haag duality breaking}\label{Haag_section}

 Orbifolds with symmetry breaking present algebra region problems even in regions with trivial topology. The breaking of Haag duality for disks is one manifestation of it. As explained in section \ref{algebrasRegions}, this can be restored at the expense of losing the additivity property. The Haag duality breaking can be tested straightforwardly in the radial lattice for the Maxwell field just by comparing the commutant algebra associated with a region with the algebra of the complement.
 If we take, for example, the subalgebra $\mathcal{C}$ in the segment $1\leq i\leq n$, where $R=n+1/2$, then the commutant algebra is 
\begin{equation}
\mathcal{C}'_R =\lbrace (\phi_n,\sum_{i=1}^n \sqrt{i}\pi_i),(\phi_{n+1},\pi_{n+1}),...,(\phi_m,\pi_m)\rbrace \label{commutant} \footnote{We remove the tildes to avoid cumbersome notation}.
\end{equation} 

\begin{figure}[t]
\begin{center}  
\includegraphics[width=0.55\textwidth]{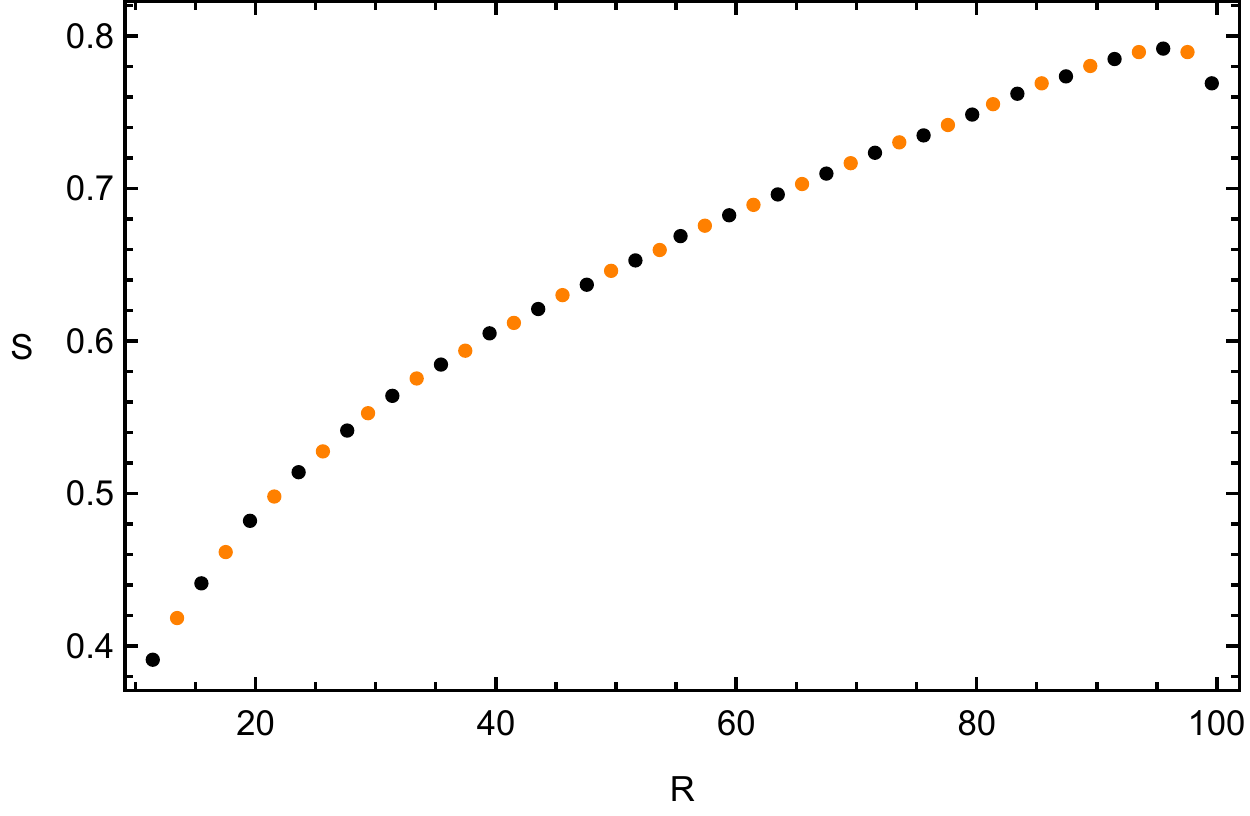}
\caption{Maxwell in a disk: Entropy vs. disk radius. The black points correspond to Maxwell subalgebra $\mathcal{C}$, whereas the yellow ones correspond to $\mathcal{C}'$. Here, $\mathcal{C}_{R'}\neq\mathcal{C'}_{R}$. Calculations are done in a lattice of size $m=100$.}
\label{haag}
\end{center}  
\end{figure}

\begin{figure}[t]
\begin{center}  
\includegraphics[width=0.55\textwidth]{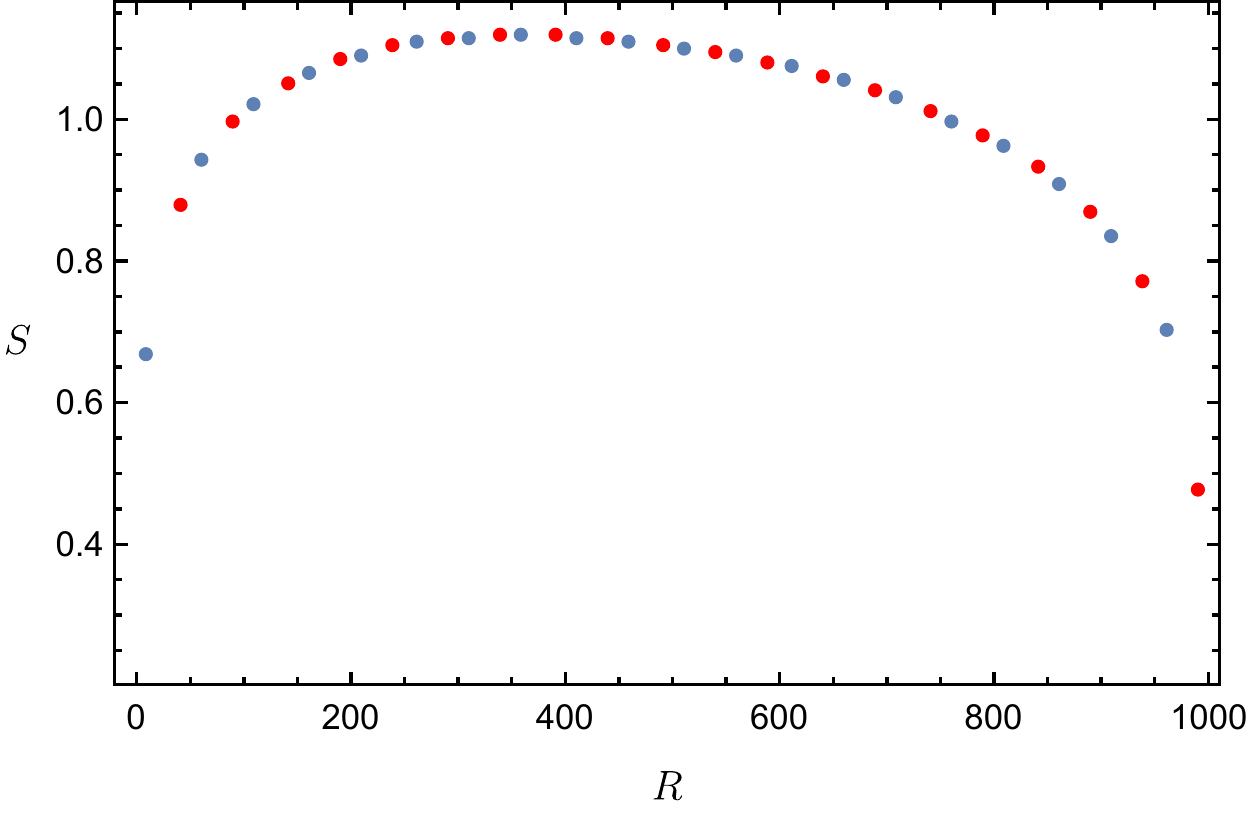}
\caption{Scalar in a disk:  Entropy vs. disk radius. The blue points correspond to $\mathcal{F}_R$ and red points to $\mathcal{F}_{R'}$ in the complementary region. Here, Haag duality is satisfied $\mathcal{F}_{R'}=\mathcal{F'}_{R}$. Calculation done in a finite lattice ($1000$ points) for $n=0$ mode.}
\label{hdscalar}
\end{center}  
\end{figure}

Namely, the commutant is made up of the full scalar algebra outside the disk, and an extra mode that contains the operator $\sum_{i=1}^n\sqrt{i}\pi_i$. Clearly, $\mathcal{C'}_R$ does not coincide with the algebra in the complementary region $\mathcal{C}_{R'}\neq \mathcal{C'}_{R}$. On the other hand, given that the entropies of pure states in commutant algebras must agree we have
\begin{equation}
S(\mathcal{C}_R)=S(\mathcal{C}'_R)=S(\mathcal{F}_{R'}\vee \lbrace(\phi_n,\tau(0\rightarrow R))\rbrace) 
\end{equation}
where we denote $\tau(0\rightarrow R)$ the  twist operator \cite{Casini:2019kex} defined in the radial lattice as
\be
\tau(n_1\rightarrow n_2)=\sum_{i=n_1}^{n_2}\sqrt{i}\pi_{i}\,.
\ee
These are non locally generated operators within the region $V'_{(n_1,n_2)}$ and are the ones that implement the symmetry.  
We numerically computed the EE both in $\mathcal{C}_R$ and $\mathcal{C}_R'$ written in (\ref{commutant}) and found a perfect match. These results are shown in figure (\ref{haag}). The equality between entropies, in turn, serves as a cross-check for the operator content of the commutant algebra $\mathcal{C}_R'$, that in cases like this, in which Haag duality is broken, may be highly nontrivial. 
On the contrary, Haag duality is satisfied for the $n=0$ mode of the scalar field in the disk, as expected for a complete theory. This is shown in figure (\ref{hdscalar}).

Since later we are going to study mutual information between a disk of radius $R$ and the complementary region  of a disk of radius $R+d$, with $d \ll R$, we complete the analysis considering the two other relevant cases: the outer region of a disk $R+d\equiv k+1/2$ and a two-component region corresponding to the union of a disk $ R \equiv n+1/2$ and the complement of a disk $ R + d $. Note that for multi-component regions, the combination of the subalgebras listed above for each component results in many more possible choices in the total algebra assignment. We present here some examples:
\begin{equation}\begin{aligned}
\mathcal{B}'_{(R+d)'}=&\mathcal{F}_{R+d}\vee \lbrace (\phi_{k+1},\tau(R+d\rightarrow \infty))\rbrace \\
\mathcal{C}'_{R\cup (R+d)'}=&\mathcal{F}_{(R,R+d)}\vee \lbrace(\phi_n,\tau(0\rightarrow R))\rbrace \vee\lbrace(\phi_{\infty},\tau(R+d\rightarrow \infty))\rbrace\hspace{-10pt}\end{aligned}
\label{hdb}
\end{equation}
where the sub-index $(R,R+d)$ denotes the shell with inner radius $R$ and outer radius $R+d$. 

Note that the twist operators appearing in the above commutant algebras are quantum operators contributing to the quantum entropy as the rest of the harmonic modes. However, if we had chosen subalgebra $\mathcal{E}$, then the twist operators would have belonged to the center, thus giving rise to a classical contribution. Stated differently, the quantum/classical character of the twists depends on the algebra assigned to the region.

\subsection{Instability of disk entropy difference}
As explained in section \ref{themodel}, we will show the difference between the disk entropy of the Maxwell theory and that of the scalar is a good order parameter for the spontaneous symmetry breaking of the orbifold model only when it corresponds in the continuum to a relative entropy. In that case, it grows logarithmically with the disk radius, with a coefficient $c_{\text{log}}=1/2$ as predicted in the literature. Aiming to numerically verify this behavior, we compute $\Delta S$ in the radial lattice, and compare the results obtained for the different algebra choices.

Once again, note that we just need to care about the $n=0$ mode, since $\Delta S=S_{n=0}^M-S_{n=0}^S$.\footnote{In order to present the results in a clearer way, in this section alone we define the entropy difference as minus that of eq.(\ref{log}).} Hence $c_{\text{log}}^M-c_{\text{log}}^S\equiv \Delta c_{\text{log}}$, with $c_{\text{log}}^S$ given in (\ref{clog_scalar}).

The coefficient $c_{\text{log}}^M$ has been calculated using the method described in section \ref{numerics}. The entropy for the subalgebras  $\mathcal{C}$,  $\mathcal{B}$ and $\mathcal{D}$ is given by eq. (\ref{vonNeumann}) and for $\mathcal{E}$ by eq (\ref{vonNeumann2}). The correlators (\ref{corM}) were obtained as the infinite lattice size limit $m \rightarrow \infty$ of the ones calculated for different sizes $m=2000,3000,4000,5000,6000$. We fitted the numerical data with the function $a_0+\frac{a_1}{m}+\frac{a_2}{m^2}$ and identified the correlator for infinite lattice size with the coefficient $a_0$. We considered disks with radius $R=5,6,...,34,35$ and fitted the data with the function $c_0+c_{\text{log}}\log t+\frac{c_{-1}}{t}$.

We list $\Delta c_{\text{log}}$ for different Maxwell subalgebras in the folowing table
\begin{table}[H]
\begin{center}
\begin{tabular}{|c|c|c|c|c|}
\hline
Algebra & $\mathcal{B}_R$ & $\mathcal{C}_R$ & $\mathcal{D}_R$ & $\mathcal{E}_R$\\
\hline
$\Delta c_{\text{log}}$ & 0.499 & $7.43\times 10^{-5}$ & $\sim 0.08$ & 0.500 \\
\hline
\end{tabular}
\end{center}
\caption{Logarithmic coefficient of the EE for Maxwell subalgebras}
\label{table1}
\end{table}

These results become highly relevant in terms of what they imply regarding the universal contribution (\ref{log}): they reveal that the entropy difference is unstable. More concretely, the topological logarithmic contribution,  well defined in the continuum through a relative entropy, is  not trustfully captured in the radial lattice through the EE. Some choices reproduce the continuous result and some do not. In other words, only for the subalgebras $\mathcal B_R$ and $\mathcal E_R$ the relative entropy involving the full scalar vacuum $\omega$ and the projected vacuum $\omega\circ E$ can be identified with $\Delta S$. 

The results for the subalgebras $\mathcal{B}_R$ and $\mathcal{C}_R$ can be accounted for by analyzing the respective commutant algebras, which are identical to that of the scalar, save an extra mode. For example, 
\begin{equation}
\mathcal{B}'_R=\lbrace (\phi_1,\sum_{i=1}^n \sqrt{i}\pi_i),(\phi_{n+1},\pi_{n+1}),...,(\phi_m,\pi_m)\rbrace \,,\label{brcommutant}
\end{equation}
whereas the commutant of the full algebra in the disk is 
\begin{equation} 
 \mathcal{F}_R'=\lbrace(\phi_{n+1},\pi_{n+1}),...,(\phi_m,\pi_m)\rbrace\,.\label{frcommutant}
\end{equation} 
 
Therefore, the difference in entropies can be thought of as stemming from the contribution of this additional mode.  For a large enough region ($n>>1$), this goes roughly as
\begin{equation}
\Delta S\sim \log{\sqrt{\langle \phi_1 \phi_1\rangle \sum_{i=1}^n\sum_{j=1}^n \sqrt{i}\sqrt{j}\langle \pi_i \pi_j \rangle}}\sim \frac{1}{2}\log{n} +\text{const.}\, ,
\end{equation}
supporting the numerical result $\Delta c_{\text{log}}\sim 1/2$ shown in the first column of table (\ref{table1}).

\begin{figure}[t]
\begin{center}  
\includegraphics[width=0.55\textwidth]{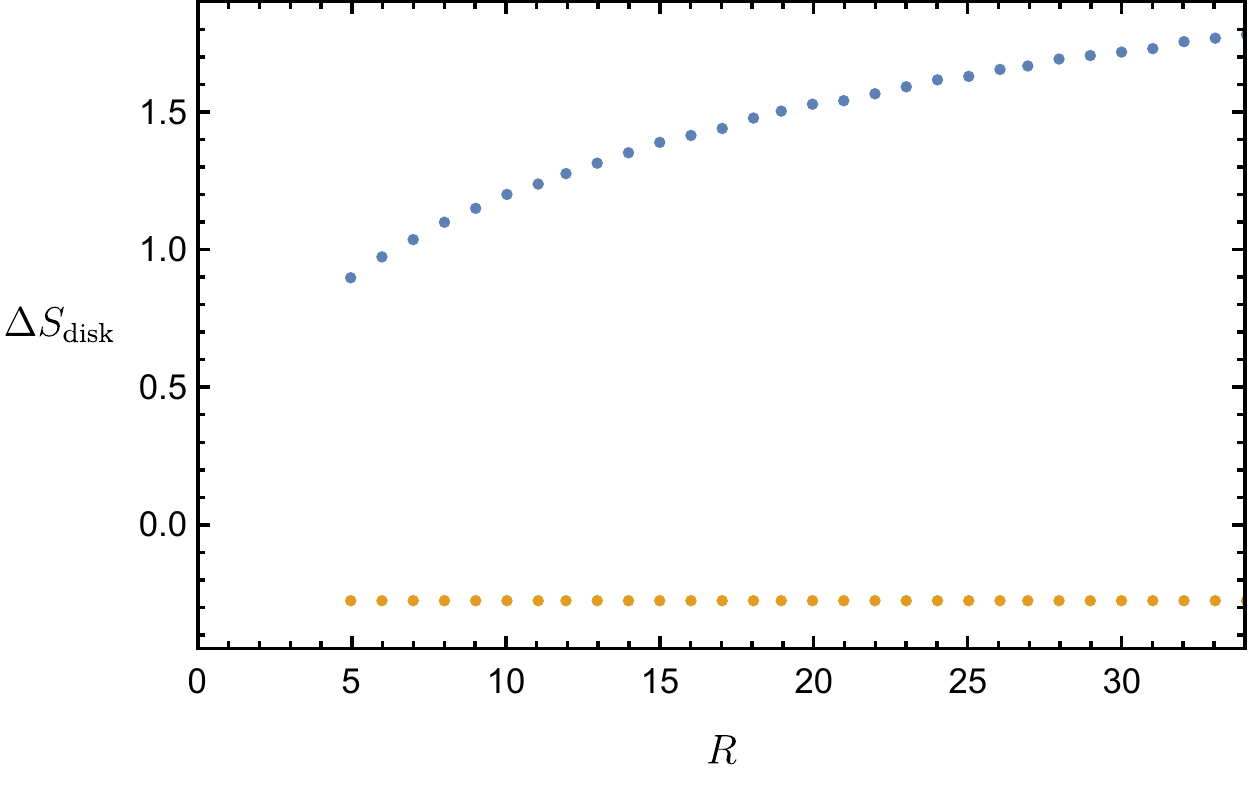}
\caption{$\Delta S$ for subalgebras $\cal{B}$(blue) and $\cal{C}$(yellow)}
\label{differences}
\end{center}  
\end{figure}

As anticipated in section (\ref{Haag_section}), $\mathcal{C}'_R$ involves instead the extra mode $(\phi_n,\sum_{i=1}^n \sqrt{i}\pi_i)$, with commutation relation $\left[\phi_n,\sum_{i=1}^n \sqrt{i}\pi_i\right]\propto \sqrt{n}$. Hence, 
\begin{equation}
\Delta S\sim \log{\sqrt{\frac{1}{n}\langle \phi_n \phi_n\rangle \sum_{i=1}^n\sum_{j=1}^n \sqrt{i}\sqrt{j}\langle \pi_i \pi_j \rangle}}\sim \text{const.}\, ,
\end{equation}
which justifies why $\Delta c_{\text{log}}\sim 0$ when subalgebra $\mathcal{C}_R$ is chosen.
$\Delta S$ for these two cases is shown in figure (\ref{differences}). 

\begin{figure}[t]
\begin{center}  
\includegraphics[width=0.55\textwidth]{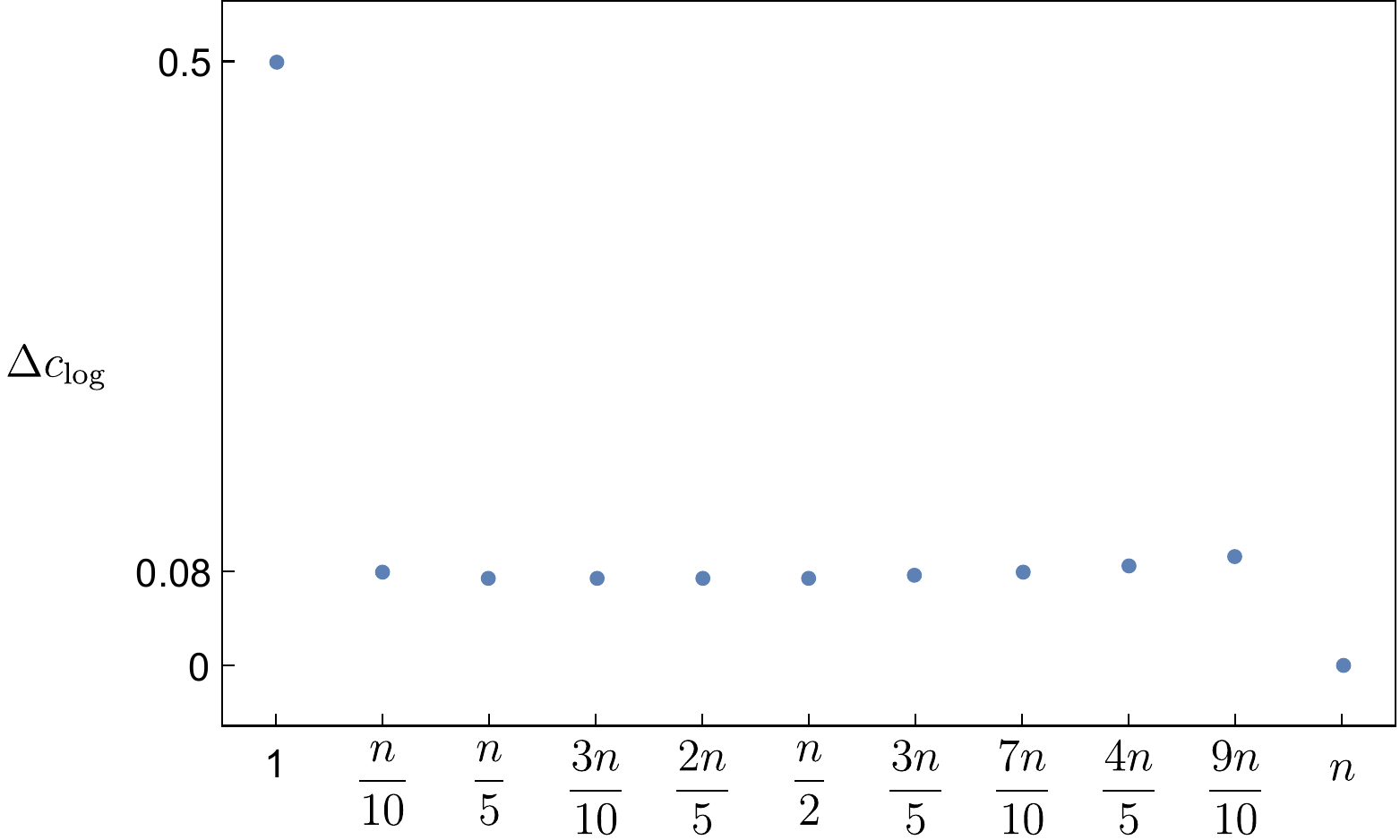}
\caption{Logarithmic coefficient of the entropy for the Maxwell field, relative to the scalar's. The horizontal axis represents the lattice site with a missing $\pi$ operator, $n$ being the total size of the region. The result is roughly independent of the position, as long as it scales with the total size.}
\label{ClogVsPi}
\end{center}
\end{figure}

On the other hand, the same argument does not apply for an algebra $\mathcal{D}_R$ with a missing $\pi_k$ operator, site $k$ being some fraction of the total size $n$. In figure (\ref{ClogVsPi}) we plot $\Delta c_{\text{log}}$ as a function of the site without momentum operator, and show that the result is approximately constant provided that the site is not at the end points. We get $\Delta c_{\text{log}}\sim 0.08$. This is not surprising, since this configuration mimics the behavior of a $2d$ scalar field EE in a region with defects \cite{peschel}, which induce a correction to the logarithmic term of precisely $\Delta c_{\text{log}}^{\text{eff}}\sim 1/12$. In fact, we have included this choice only for completeness reasons, since naturally the introduction of defects in the region leads us to an absolutely different problem. 

Finally, as long as the algebra $\mathcal{E}_R$ (defined in  section \ref{Maxwell algebras in the radial lattice}) is concerned, $\mathcal{E}'_R$ is different from the scalar due to the presence of a center containing the twist $\tau_{1\rightarrow n}=\sum_{i=1}^n \sqrt{i}\pi_i$. Its contribution is given by the classical entropy (\ref{Shannon}),
\begin{equation}
\Delta S\sim \frac{1}{2}\log{\sum_{i=1}^n\sum_{j=1}^n \sqrt{i}\sqrt{j}\langle\pi_i\pi_j\rangle}+\text{const}\sim \frac{1}{2}\log{n}+\text{const},
\end{equation}   
which is consistent with the numerical result $\Delta c_{\text{log}}\sim 1/2$ in the last column of table (\ref{table1}).

As discussed in section \ref{themodel}, the universality of $\Delta S$ is based on the identification $\Delta S = S_{\mathcal{ F}}(\omega\vert\omega\circ E)$. In fact, when the choice of the local algebra is such that the respective entropies do not satisfy $S_{\mathcal{O}}(\omega) \longleftrightarrow S_{\mathcal{F}}(\omega \circ E)$, then  $\Delta S \neq S_{\mathcal{F}}(\omega)-S_{\mathcal{ F}}(\omega\circ E)=S_{\mathcal{ F}}(\omega\vert\omega\circ E)$ and consequently equation (\ref{log}) is violated.  Hence, it is clear that the choice ${\cal B}_R$ gives rise to an entropy $S_{\mathcal{O}}(\omega)$ which cannot be identified with $S_{\mathcal{F}}(\omega \circ E)$, as opposed to the choices ${\cal C}_R$ and ${\cal E}_R$.

In this regard, it is important to note that in the ${\cal C}_R$ case (\ref{cr}),  the twist operator $\tau_{1\rightarrow n}=\sum_{i=1}^n \sqrt{i}\pi_i$, that implements the symmetry within the disk, does not belong to the disk algebra due to the missing $\pi_n$ at the boundary, which in turn, causes the twist to appear in the commutant as an extra quantum mode $(\phi_n,\tau_{1\rightarrow n})$ (\ref{commutant}). This leaves us with an outer conditional expectation $E_{\text{out}}$ as opposed to an inner one $E_{\text{inn}}$, implemented by operators belonging to the algebra of the disk and for which the condition $S_{\mathcal{O}}(\omega) \longleftrightarrow S_{\mathcal{F}}(\omega \circ E_{\text{inn}})$ is guaranteed.
On the other hand, for the choices ${\cal E}_R$ (\ref{er}) and ${\cal B}_R$ (\ref{br}), the twist  $\tau_{1\rightarrow n}$ does belong to the disk algebra, explicitly in the first case, and effectively in the second, both with an inner conditional expectation $E_{\text{inn}}$ associated. When the ${\cal B}_R$ realization is chosen, the twist appears in the commutant as an extra gaussian mode $(\phi_1,\tau_{1\rightarrow n})$ (\ref{brcommutant}), but this time, due to the missing $\pi_1$ operator, independently of the disk size (contrary to what happens for the ${\cal C}_R$ choice). Noting that the missing $\pi_1$ or in the dual commutant picture, the extra  $\phi_1$, represent operators living in a ball of size $\delta$ placed at the center of the disk, they become irrelevant once the continuum limit is taken, promoting the twist as an effective center with an associated $E_{\text{inn}}$.

\subsection{Mutual information difference between $\mathcal{F}$ and $\mathcal{O}$}

In this section we test eq. (\ref{loglog}) numerically. As an order parameter for duality/additivity breaking due to the presence of non-local operators, $\Delta I$ is a relative entropy associated to regions with non-trivial topology. Being a well defined quantity in the continuum limit, free of divergences and independent of the regularization scheme, the universal terms that come about in the radial lattice calculation become stable, unaware of the regularization issues that we discussed above for the entropy.

As explained in \ref{themodel}, we are interested in the mutual information between a disk of radius $R$ and the complement of a disk of radius $R+d$ in the limit $d\rightarrow 0$, corresponding to nearly complementary regions. This is
\begin{equation}\label{mutual_3}
I_{M,S}=S_{M,S}(R)+S_{M,S}((R+d)')-S_{M,S}(R \cup (R+d)').
\end{equation}
Here the subindices $M$ and $S$ denote Maxwell and scalar algebras, respectively. The second and third term involve unbounded regions that we are going to treat as usual, considering instead the commutant algebras associated conveniently to finite regions.
 Furthermore, given that we will ultimately compute the difference between the Maxwell mutual information and the scalar counterpart, it is more insightful to work with the commutant algebras of each region, which differ only in the appearance of twist operators, as discussed in the previous section. Although in the scalar theory the commutant algebras trivially correspond to the operators in the complementary region, thanks to the Haag duality, we know that this is not the case when we consider instead the commutants in the Maxwell subalgebra.
 
We will analyze the mutual information for a particular algebra choice and discuss the differences that might arise if a different assignation is made (remember that there are multiple possibilities by combining different options for each region's component). We note it is not necessary to explore all possible realizations, since due to the regularization scheme independence of the MI, it is evident that any other combination will result in the same $\Delta I$.

Let us define $R=n+1/2$ and $R+d=k+1/2$ and choose $\mathcal{B}_R$ and $\mathcal{B}_{(R+d)'}$ the subalgebras assigned to the disk of radius $R$ and the complement of a disk of radius $R+d$ respectively. The corresponding commutants are given by
\begin{equation}
\mathcal{B}'_R =\lbrace (\phi_1,\sum_{i=1}^n \sqrt{i}\pi_i),(\phi_{n+1},\pi_{n+1}),...,(\phi_m,\pi_m)\rbrace = \lbrace (\phi_1,\sum_{i=1}^n \sqrt{i}\pi_i)\rbrace	\vee \mathcal{F}_{R'}\,,
\end{equation}
\begin{equation}
\mathcal{B}'_{(R+d)'}= \lbrace (\phi_1,\pi_1),...,(\phi_k,\pi_k),(\phi_{k+1},\sum_{i=k+1}^m \sqrt{i}\pi_i)\rbrace = \mathcal{F}_{R+d} \vee \lbrace (\phi_{k+1},\sum_{i=k+1}^m \sqrt{i}\pi_i)\rbrace\,,
\label{commutantbr+d}
\end{equation}
\begin{equation}
\begin{split}
\mathcal{B}'_{R \cup(R+d)'}&= \lbrace (\phi_1,\sum_{i=1}^n \sqrt{i}\pi_i),(\phi_{n+1},\pi_{n+1}),...,(\phi_{k},\pi_{k}),(\phi_{k+1},\sum_{k+1}^m\sqrt{i}\pi_i)\rbrace\\
&=\lbrace (\phi_1,\sum_{i=1}^n \sqrt{i}\pi_i)\rbrace \vee \mathcal{F}_{(R, R+d)}\vee \lbrace (\phi_{k+1},\sum_{k+1}^m\sqrt{i}\pi_i)\rbrace\,,
\end{split}
\label{commutantbr+dunionr}
\end{equation}
where we have included in (\ref{commutantbr+dunionr}) the commutant of the subalgebra of the union $\mathcal{B}'_{R \cup(R+d)'}$ present in the mutual information definition. 
For the sake of clarity, an outline of the algebras involved in the mutual information is shown in figure (\ref{DeltaI_algebras}).

The first piece in (\ref{mutual_3}) follows straightforwardly from the analysis made in the previous section. In the present case, the additional mode is responsible for the behavior\footnote{Beware that the definition adopted in the previous section for the difference $\Delta S$ is minus the one used here.}
\begin{equation}\label{S_disk}
\Delta S_{\text{disk}}\equiv S_S(R)-S_M(R)=S(\mathcal{F}_R)-S(\mathcal{B}_R)\sim -\frac{1}{2}\log{R}\,.
\end{equation}
The same would have applied if we had chosen $\mathcal{E}_R$ (see table (\ref{table1})). On the contrary, $\mathcal{C}_R$ would have lead to no logarithmic contribution. 

Meanwhile, the presence of unbounded twist operators makes the computation of the second and third terms in (\ref{mutual_3}) more costly. However, we will argue that once the infrared and the continuum limits are taken (in that order), their contribution can be neglected in the calculation.  The unbounded twist is present both in the second and third terms with opposite sign. Its contribution to the entanglement as we take the infinite lattice size limit decouples from the rest, allowing the cancellation. Being a mode with divergent self correlation in the infinite lattice limit $\langle\tau,\tau\rangle\rightarrow\infty$, it will decouple from the rest of the system (in accordance with the monogamy property of the entropy  \cite{petz2007quantum,araki1976relative}). Let us see this in more detail. 

Consider the case of two regions: $V_1$, a disk of radius $R$, and $V_2$, a shell $(R-d,R)$. In addition, consider some algebras $\mathcal{A}_{1,2}$ made up of the operators locally generated in $V_{1,2}$, together with the unbounded twist $\tau_{R,\infty}$. As discussed in section \ref{Haag_section}, this may or may not be part of the classical center, depending on our algebra choice. 

Let us analyze first the quantum case, such that $\mathcal{A}_i=\mathcal{A}_{V_i}\vee \{(\phi_R,\tau_{R,\infty})\}$ with $i=1,2$.
It is possible to see that the difference  $S(\mathcal{A}_1)- S(\mathcal{A}_2)$ can be replaced by $S(\mathcal{A}_{V_1})- S(\mathcal{A}_{V_2})$ since
\begin{eqnarray}
S(\mathcal{A}_{V_1}\vee \{(\phi_R,\tau_{R,m})\})- S(\mathcal{A}_{V_2}\vee \{(\phi_R,\tau_{R,m})\})&\xrightarrow[m \to \infty]{}& S_Q(\mathcal{A}_{V_1})+H(\phi_{R})-(S_Q(\mathcal{A}_{V_2})+H(\phi_{R}))\nonumber\\
&\xrightarrow[\epsilon \to 0]{}& S(\mathcal{A}_{V_1})- S(\mathcal{A}_{V_2})\,.
\label{infqtwist}
\end{eqnarray}

An explanation is in order. Firstly, we take the infrared limit, which, as we already argued, enables us to get rid of the unbounded twist operator. What results from this operation  is an algebra with center: the operator $\phi_R$ does not commute
with the twist, that contains in the sum the momentum $\pi_R$, but once the twist is eliminated it becomes a center of the algebra. For algebras with center we know the entropy is composed by two parts, the quantum entropy $S_Q$, and $H(\phi_R)$, the classical contribution of the center. Secondly, we take the continuum limit, in which we can safely neglect the contribution of the operator $\phi_R$ to the quantum piece and $\Delta S_Q\rightarrow \Delta S$. 
This is related to the fact that in the continuum limit the classical contributions properly combined in the mutual information vanish \cite{Casini:2014aia} and the quantum part  becomes independent of the presence of the center. In fact, equation (\ref{infqtwist}) could be completed to obtain a mutual information and the above discussion would apply directly.
We have checked (\ref{infqtwist}) numerically, getting perfect agreement for $\langle\tau,\tau\rangle\sim 1000$.
Note that the order of limits is very important.
Equation (\ref{infqtwist}) will be useful in the analysis for the subalgebras $\mathcal{B}$ and $\mathcal{C}$, both involving an unbounded quantum twist.

If, instead, the twist operator forms a center, as in the subalgebra $\mathcal{E}$, the analysis is even easier:
\begin{figure}[t]
\begin{center}  
\includegraphics[width=0.9\textwidth]{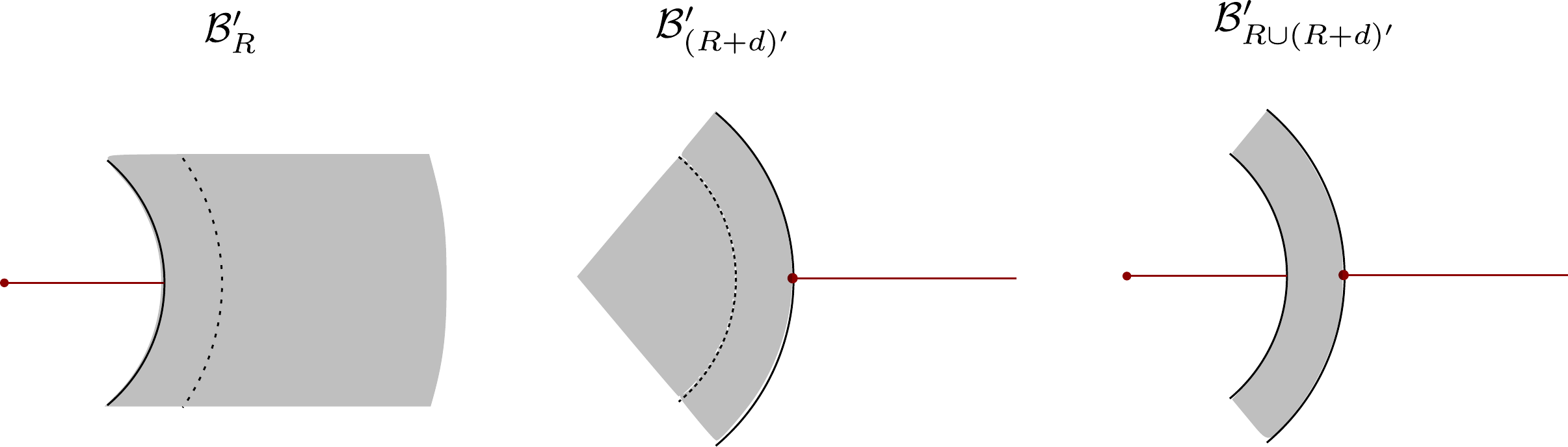}
\caption{Commutant of Maxwell subalgebras. Grey regions represent the scalar algebra, while the red segments depict the twist operator, linked to a single scalar mode at the dot.}
\label{DeltaI_algebras}
\end{center}
\end{figure}
\begin{eqnarray}
S(\mathcal{A}_{V_1}\vee\tau_{R, \infty})- S(\mathcal{A}_{V_2}\vee\tau_{R, \infty})&=& S_Q(\mathcal{A}_{V_1})+H(\tau_{R,\infty})-(S_{Q}(\mathcal{A}_{V_2})+H(\tau_{R,\infty})) \nonumber \\
&\xrightarrow[\epsilon \to 0]{}& S(\mathcal{A}_{V_1})- S(\mathcal{A}_{V_2})
\label{infcltwist}
\end{eqnarray}
Here, we start with algebras with center in the limit $m \rightarrow \infty$. The first equality is simply the application of the definition of the entropy for algebras with non trivial center. The last step is again inspired on the universality of the MI with respect to the presence of centers in the continuum limit, and was numerically checked.

Resuming the analysis for the choice $\mathcal{B}_R$ and $\mathcal{B}_{(R+d)'}$, according to (\ref{commutantbr+d}) and (\ref{commutantbr+dunionr}), and considering (\ref{infqtwist}),  we get 
\begin{equation}
S_M((R+d)')- S_M(R \cup (R+d)')=S(\mathcal{F}_{(R+d)})-S(\mathcal{G}),
\end{equation}
where
\begin{equation}
\mathcal{G}\equiv\lbrace (\phi_1,\sum_{i=1}^n\sqrt{i}\pi_i)\rbrace \vee \mathcal{F}_{(R,R+d)}
\end{equation}
with commutant $\mathcal{G}'$
\begin{equation}
\mathcal{G}'= \mathcal{B}_R \vee \mathcal{F}_{(R+d)'}\,.
\end{equation}
With these elements in place, the remaining contribution to $\Delta I$ is given by the difference between the entropy of the scalar field at the shell and that of the full algebra at the shell, plus a twist operator from the origin to $R$. That is, 
\begin{equation}
\Delta S_{\text{shell}}\equiv \Delta S((R+d)')- \Delta S(R \cup (R+d)')	= S(\mathcal{G})-S(\mathcal{F}_{(R,R+d)})
\end{equation}

\begin{figure}[t]
\begin{center}  
\includegraphics[width=0.7\textwidth]{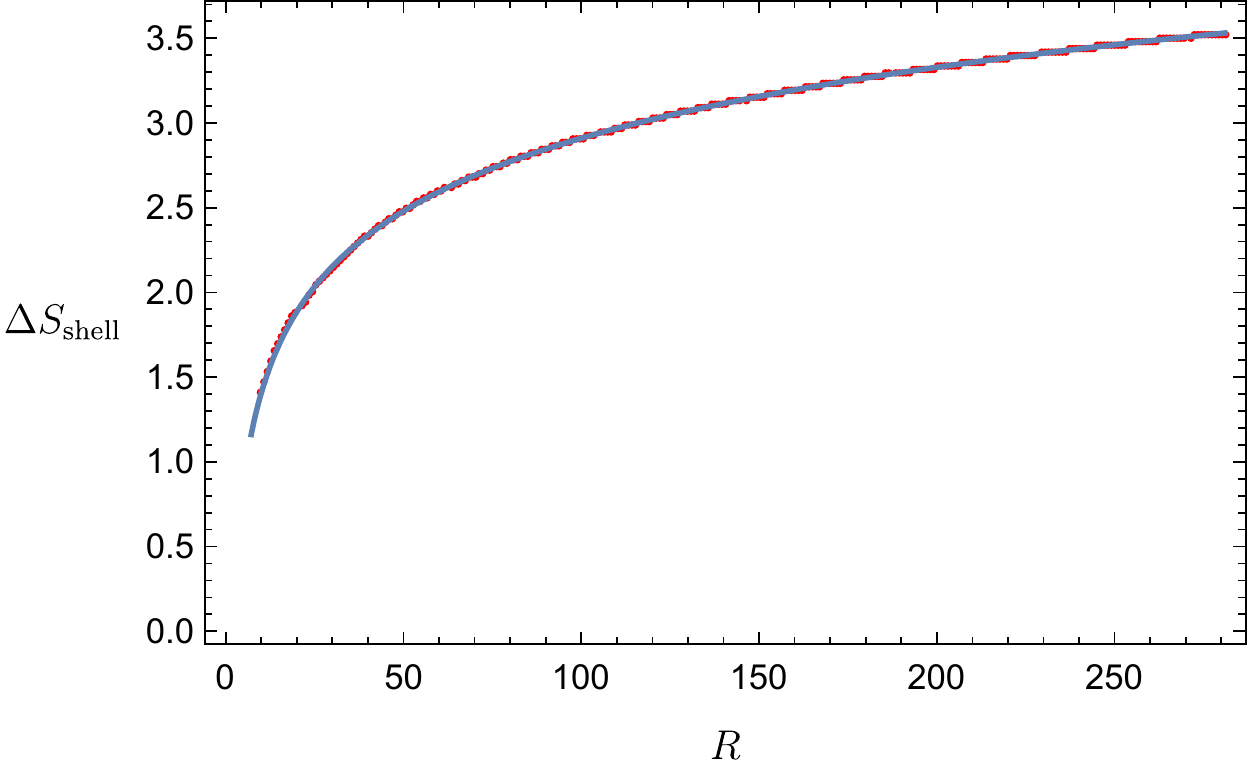}
\caption{$\Delta S_{\text{shell}}$ vs. disk radius $R$. The red points are the numerical values, and the blue curve represents the best fit $f(R)=-0.169296+0.502607\log{R}+0.500937\log{\log{R}}$. A total lattice size of $m=1000$ was used, but the correlators were obtained from the $m=5000$ radial lattice to better approximate the exact discrete correlators.}
\label{mutual_shell}
\end{center}
\end{figure}
Setting the shell width to the unit radial lattice size, numerical calculations yield	
\begin{equation}\label{S_shell}
\Delta S_{\text{shell}}\sim \frac{1}{2}\log{R}+\frac{1}{2}\log{\log{R}}.
\end{equation} 
This agrees with the analytic approximation of the twist algebra entropy reported in \cite{Casini:2019kex}. See figure (\ref{mutual_shell}) for a more detailed account on the numerical results.

Note that the logarithmic contribution of (\ref{S_disk}) cancels that of (\ref{S_shell}), leading to
\begin{equation}
\Delta I= \Delta S_{\text{disk}}+\Delta S_{\text{shell}}\sim \frac{1}{2}\log{\log{R}},
\end{equation}
as expected.  

It is interesting to note that, no matter the algebra choice, the difference $\Delta I$ can always be expressed in terms of differences between disk and shell entropies. 

Moreover, it is worth stressing that had we chosen, for example, Maxwell subalgebra $\mathcal{C}_R$ and $\mathcal{C}_{(R+d)'}$, then both (\ref{S_disk}) and (\ref{S_shell}) would have changed in their logarithmic contribution, but the two logarithmic pieces would cancel anyway. We have numerically calculated (\ref{S_shell}) for this subalgebra choice  and obtained
\begin{equation}\label{S_shell2}
\Delta S_{\text{shell}}\sim \frac{1}{2}\log{\log{R}}.
\end{equation} 
which together with $\Delta S_{\text{disk}}\sim 0$ consistently leads to the same $\Delta I$ as above.

Finally, for the case $\mathcal E_R$ and $\mathcal{E}_{(R+d)'}$, thanks to (\ref{infcltwist}) we can follow the same steps as before and evaluate $\Delta S_{\text{disk}}+\Delta S_{\text{shell}}$ without taking into account the unbounded twists. According to the discussion in the previous section, the entropy difference is $\Delta S_{\text{disk}}\sim -1/2 \log R$. Meanwhile, $\Delta S_{\text{shell}}$ can be calculated analytically \cite{Casini:2019kex} and gives the same result as that associated to subalgebra $\mathcal{B}$, so $\Delta I$ remains unaffected. 

As we mentioned at the beginning,  the same $\log \log$ contribution is guaranteed for all possible assignations due to the regularization scheme independence of the MI. This legitimates mutual information as a well defined information measure and a good order parameter for the model incompleteness.

Furthermore, note that for the Maxwell field the mutual information of nearly complementary spherical regions is not a regularized version of the disk entanglement entropy, as can be deduced by comparing equations (\ref{log}) and (\ref{loglog}). If we had obtained the same logarithmic behavior as the entropy (present for the particular algebra choices for which it becomes a relative entropy), we would have arrived at a violation of monotonicity. What prevents this from happening is the contribution of the non local operators responsible for the incompleteness of the model. More concretely, the entropy of the union does not vanish as usual in the limit of small separation, but rather diverges, as shown in (\ref{S_shell}). And that contribution is exactly that of the twist algebra implementing the symmetry in the disk.

\section{Final remarks}

We compute numerically the EE of the Maxwell field for different rotationally invariant regions with the purpose of checking the novel topological contributions to the EE and MI due to the "incompleteness" of the model, predicted in \cite{Casini:2019kex,Casini:2020rgj}. In this setup,  it turns out that once the problem is reduced to the half-line,  thanks to the rotational symmetry, the subalgebra of the derivatives of the scalar field is equivalent to the one of the fields themselves except for the Fourier $n= 0$ mode $\phi_0$. This is reminiscent of what happens in $d=4$ for the sphere. In four dimensions, the Maxwell and scalar theories also differ in the zero mode, only just that the $\phi_0$ completely disappears from the Maxwell theory. 

Here, we explore different lattice realizations for disks $D_R$ and find that only some of them give the expected $\Delta S\sim -\frac{1}{2}\log(R/\delta)$. As suggested in \cite{Casini:2019kex,Casini:2020rgj}, the entropy difference is unstable and depends on the lattice details. The universal character of the logarithmic correction for the entropy relies on the identification of the difference $\Delta S$ with a relative entropy, well defined in the continuum. We show explicit realizations where this identification fails causing, in turn, the failure of the universal character of the correction.

Contrary to the EE, the mutual information difference is for all the subalgebra choices in perfect agreement with the predicted result. The reason is clear, being not only the difference but each mutual information itself a relative entropy.
 
Following the same line of reasoning, we can also understand why mutual information and entropies depend differently on $R$. In incomplete models with spontaneous broken symmetry, it is not possible to define a regularized entropy through the mutual information. That is, $2S^{reg}_R=I(R^-,{R^{+}}')$, in the nearly complementary regions limit $R^- \sim R^{+}$. The reason is that this identification relies on the Haag duality property, which is not satisfied for some subalgebras. As explained in the last section, it is the emergence of extra non-local operators that spoils the interpretation of the MI  as a regularized entropy and in turn disconnects the universal character of both quantities.

We also find interesting issues related to unbounded twist operators. When we deal with unbounded regions, it is standard to consider instead the commutant algebra associated to the complementary bounded region, profiting that for pure states the algebra in a region and its commutant have the same EE.  In the cases presented here, the commutant algebra contains unbounded twist operators. We show that the infrared limit (infinite lattice size) followed by the continuum limit (zero lattice spacing), necessary to extract a quantity of the continuum, removes the unbounded twists no matter if we have chosen a subalgebra with quantum or classical twists. This cancellation occurs in MI where the same unbounded twist appears in one of the regions and the union, with opposite signs.

\section*{Acknowledgements}
We thank Horacio Casini and Diego Pontello for the enriching discussions while this work was being carried out. This work was supported by CONICET, CNEA and Universidad Nacional de Cuyo, Instituto Balseiro, Argentina.




\bibliography{refs}

\bibliographystyle{utphys}

\end{document}